\documentclass[journal]{IEEEtran}
\usepackage{cite}
\usepackage{times,amsmath,amssymb,psfrag}
\usepackage{graphicx}
\usepackage{url}
\usepackage{bm}
\usepackage{multicol}
\graphicspath{{./pics/}} \DeclareGraphicsExtensions{.pdf,.jpg,.png}
\usepackage{mdframed}
\usepackage{xcolor}
\usepackage{hyperref}
\usepackage{epsfig}
\usepackage{float}
\usepackage{array}
\usepackage{multirow}
\usepackage{etoolbox}
\usepackage{nomencl}
\usepackage{algorithm}
\usepackage{algorithmicx}
\usepackage{algpseudocode}
\usepackage{amsmath}
\usepackage{caption}
\usepackage{graphicx, subfig}
\usepackage{booktabs}

\ifCLASSINFOpdf
\else
\fi

\begin{document}

\title{Optimal Sensor Placement in Lithium-Ion Battery Pack for Fault Detection and Isolation}

\author{Ye~Cheng,~\IEEEmembership{Student Member,~IEEE,}\     Matilde~D'Arpino,~\IEEEmembership{Member,~IEEE,} \
        Giorgio~Rizzoni,~\IEEEmembership{Fellow,~IEEE,}
\thanks{This work has been submitted to the IEEE for possible publication. Copyright may be transferred without notice, after which this version may no longer be accessible.
\par This work is supported by NASA ULI program Electric Propulsion: Challenges and Opportunities under Grant NNX17AJ92A. (Corresponding author: Ye Cheng)
\par Y. Cheng, M. D'Arpino and G. Rizzoni are with the Center for Automotive Research, The Ohio State University, Columbus, OH 43212 USA. Y. Cheng, and G. Rizzoni are also with the Department
of Electrical and Computer Engineering, The Ohio State University, Columbus, OH, 43210 USA. (e-mail: cheng.1316@osu.edu; darpino.2@osu.edu; rizzoni.1@osu.edu).}}

\markboth{IEEE ,~Vol.~XX, No.~XX, JANUARY~2021}%
{Shell \MakeLowercase{\textit{et al.}}: Bare Demo of IEEEtran.cls for IEEE Journals}

\maketitle
\begin{abstract}
Energy storage systems for transportation and grid applications, and in the future for aeronautical applications, require the ability of providing accurate diagnosis to insure system availability and reliability. In such applications, battery packs may consist of hundreds or thousands of interconnected cells, and of the associated electrical/electronic hardware. This paper presents a systematic methodology for approaching some aspects of the design of battery packs, and in particular understanding the degree of analytical  redundancy (AR) in the system that can be used for diagnostic strategies. First, the degree of AR that is intrinsic in the battery system is determined. Then, structural analysis tools are used to  study how different measurements (current, voltage, and temperature) may improve the ability of monitoring and diagnosis of a battery system. Possible sensor placement strategies that would enable the diagnosis of individual sensor faults and individual cell faults for different battery pack topologies are analyzed as well. The work presented in this paper illustrates how to achieve the required fault detection and isolation (FDI) capabilities using a minimal or optimal sensor set, which is a critical step in the design of a large battery pack.
\end{abstract}

\begin{IEEEkeywords}
Energy storage system, Lithium-ion battery, sensor placement, fault diagnosis, structural analysis
\end{IEEEkeywords}

%
\IEEEpeerreviewmaketitle

\section{Introduction}

\begin{figure}[]
\centering
\includegraphics[width=8.5cm]{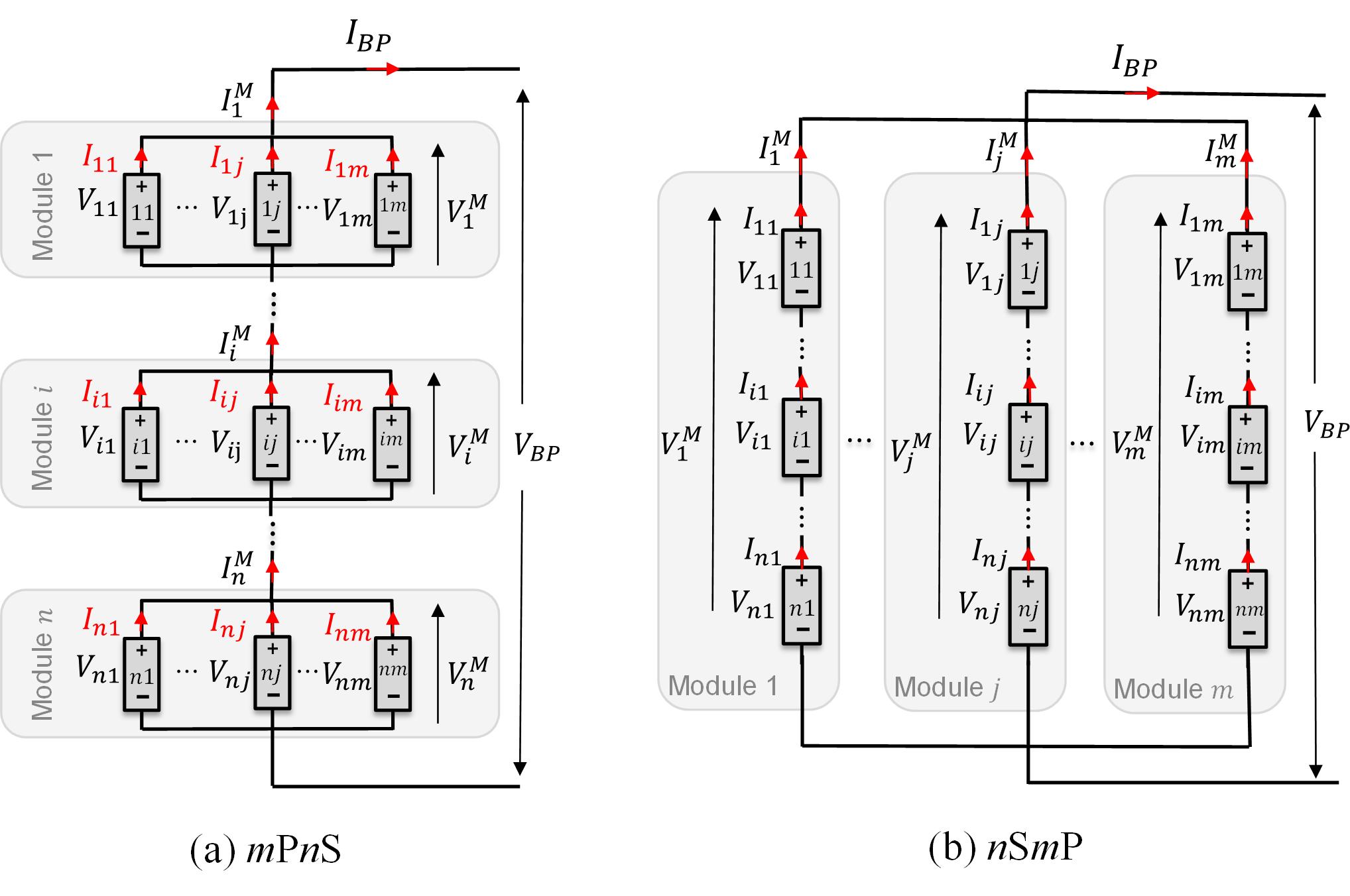}
\caption{(\textit{a}) Parallel-Series topology (PS or $m$P$n$S), (\textit{b}) Series-Parallel topology (SP or $n$S$m$P) }
\label{twotopologies}
\end{figure}

\IEEEPARstart{A}{MONG} the energy storage technologies, lithium-ion batteries (LIB) have demonstrated great capability in improving system efficiency, emissions, management of uncontrollable sources (e.g. renewable resources, regenerative braking),  controllability and power quality, system level reliability, delay system expansion/investments, weight, flexibility and modularity in several energy applications. 
Major automotive companies around the world are researching and launching electric vehicles \cite{chemali2016electrochemical}. The aircraft industry and federal agencies, such as NASA, have also invested in the research on electrified aircraft that can transport both people and cargo \cite{tariq2016aircraft,jansen2017overview, perullo2019sizing}. Similarly, electric utilities are seeking to use energy storage as a cost-effective way of supporting renewable power production and distribution \cite{faisal2018review, abronzini2019cost}. The integration of LIB in a system usually requires that battery cells are connected in series and/or in parallel to form modules, which then are assembled into battery packs to meet the energy and power requirements of vehicles or grid applications, resulting in systems that are large-dimensional and that have complex interconnections  \cite{Cai2019}. One of the open problems is the ability to properly monitor the operation of such complex systems, and to diagnose their health. When designing a large battery pack, two fundamental topologies are commonly used and according to the standard IEC 60050, they are named as parallel-series (PS), and series-parallel (SP), as shown in Fig.\ref{twotopologies}, where $i$ is the series index and $j$ is the parallel index. A battery pack is composed of $n\times m$ cells, where $n$ indicates the number of elements in series and $m$ the number of elements in parallel. The behavior of a battery pack cannot be modeled by understanding the behavior of a single cell, as the complex interconnections of cells and modules causes interactions that may limit the system performance. Because of differences in cell electrical and thermal characteristics and in cell aging, the energy/power density and the durability and safety of the battery packs will be reduced to a certain extent compared with individual cell \cite{cordoba2015control}, \cite{shu2020reliability}.  It is therefore very important to understand the behavior of large battery pack systems, which are defined by their electrical topology, by their cooling system architecture, and by the design of their battery management system. Among them, efficient sensing and fault tolerant design are important elements in the design of a battery pack. In this paper we focus on one particular aspect of the battery pack system design: the ability to diagnose faults and failures. 
\par Methods for fault diagnosis for lithium-ion battery systems can be classified into model-based, knowledge-based, and data-driven ones \cite{hu2020advanced}. The most widely used knowledge-based methods include graph theory-based (fault tree analysis) \cite{held2016safe}, expert system \cite{filippetti1992development}, and fuzzy logic-based \cite{muddappa2014electrochemical}. These diagnostic methods employ the basic knowledge and real-time observation of the battery system. Although the principle is easy to understand, before the fault diagnosis decision is made, further research is needed on the fault mechanism, knowledge acquisition and knowledge representation. Data-driven methods include signal processing \cite{zheng2013lithium,dubarry2014cell,xia2017correlation,kong2018fault}, machine learning \cite{yang2018fractional,zhao2017fault,kim2018cloud}, and information fusion \cite{huber2016method}. The advantage of these methods is that they can directly analyze and process operating data to detect failures without relying on models. The limitation in these methods is the need for large amounts of historical data, high computational costs, and training complexity. Model based methods can be divided into three categories, including  state estimation \cite{gao2018micro,sidhu2014adaptive,khalid2015health,feng2018detecting,dey2017model}, parameter estimation \cite{feng2016online,ouyang2015internal,seo2017detection,zhang2016online}, and structural analysis theory\cite{liu2014fault,liu2015structural,liu2016structural}. The development of battery models, including electrical models, thermal models and electrochemical models, provides the basis for model-based fault diagnosis. Due to the deeper insights afforded by physically based methods, these approaches can not only detect faults, but also locate faults and estimate their size. It should be noted that all of these model-based methods may be affected by model uncertainty, interference and noise.
\par Although much work has been done in fault diagnosis for lithium-ion battery, little can be found in the literature covering the topic of redundancy exploration for a battery or a battery pack. Most fault diagnosis schemes that have been proposed are based on the assumption that there are sufficient sensors providing cell information including current, voltage and temperature. For instance, in \cite{xiong2019sensor}, a sensor fault diagnosis method is presented for a lithium-ion battery pack in electric vehicles (two cells in series, which is in fact not a realistic configuration for a battery pack). This highly simplified battery pack is equipped with a voltage sensor and a temperature sensor for each cell and a current sensor. While the simplicity of this configuration with this sensor set permits experimental studies, it must be recognized that it is not practical in industrial applications that might see hundreds or thousands of cells in a battery pack. In this paper, the problem of a minimal (or otherwise optimal) sensor selection to guarantee effective FDI in a realistic battery pack is addressed.
\par A traditional approach to achieving fault tolerance may include two steps: 1) detect specific faults through limit checking or some other form of signal analysis; 2) if a failure is detected, a controller or management system will use existing redundancy to replace the faulty component or function \cite{Blanke2006}.  This approach to fault tolerance is based on \textit{physical redundancy}, where critical elements of the system are realized multiple times, with attendant increases in both system cost and complexity. However, this approach may not be always feasible in large-scale systems. In this paper, the methodology used for fault tolerance is based on understanding and exploiting system \textit{analytical redundancy} (AR) using a graphical approach known as \textquotedblleft structural analysis\textquotedblright \cite{Blanke2006,dulmage1958coverings,dulmage1959structure,krysander2008sensor}. Structural analysis is based on the systematic study of the AR inherent in a mathematical model of the system, and is especially suitable for large and nonlinear systems because it is founded on structural system properties. The system structural model is represented by a bipartite graph (usually visualized through an incidence matrix), and permits studying the AR inherent in the system with the aim of identifying FDI strategies in a systematic way. One of the outcomes of this approach is that it allows evaluation of the diagnosability of the system as a function of employing different sensor sets, and also to assist in sensor placement \cite{krysander2008sensor,meyer1994optimal,carpentier1997criteria}. 
\par The method of using structural analysis to study the FDI attributes of the system has been applied to several fields, for example automotive systems\cite{svard2010residual,sundstrom2013selecting,zhang2018fault}, engines\cite{krysander2007efficient},\cite{frisk2017toolbox}, fuel cells\cite{polverino2017model}, transmissions\cite{Chen2014SensorPA,trask2017system,ahmed2016sensors,chen2016design}, anti-lock brakes\cite{Chen2018ModelBasedFD}, drive systems of electric vehicles,\cite{Zhang2017FaultDF}, and pneumatic systems \cite{rahman2019sensor}.
The idea of applying structural analysis for battery diagnosis is not novel \cite{liu2014fault,liu2015structural,liu2016structural}, but while the results available in literature are useful and interesting, they are quite limited and not generalized to large battery packs, system faults, and sensor placement.

The contributions made in this paper may be summarized as follows: i) for the first time the tools of structural analysis are applied in a general and systematic way to a battery pack to understand the intrinsic degree of AR contained in mathematical models of the battery cells assembled in packs with two different topologies; ii) we develop general models, which include fault models, to determine the intrinsic AR of the system, in the absence of measurements; iii) the structural models are used to assess the degree of diagnosability that can be achieved for each topology by incorporating  sensors in the battery pack design.
In summary, the analysis conducted in this paper and the methods developed in it permit evaluating trade-offs among different sensor placement strategies for the purpose of diagnosis. The principal contribution of this paper is a systematic methodology to understand the diagnostic implications of sensor selections in battery packs.  While the complete design of a battery pack, including sensor selection to enable battery electrical, thermal and health management, is a complex process that involves many other design aspects, nonetheless we believe  that the work presented in this paper is an important step in this direction.

This paper is organized as follows. Section \ref{sec strcutural model} presents both mathematical model and structural model of battery. In section \ref{sec structual analysis}, the tools of structural analysis for diagnosis are introduced for battery. Section \ref{sec sensor placement} discusses the sensors placement for faults detectability and isolability of different pack topologies. Section \ref{sec comments and remarks} reports comments and remarks. Finally conclusion is drawn in Section \ref{sec concl}.

\section{Introduction to structural model of battery}
\label{sec strcutural model}

\subsection{Battery pack modeling}
The subject of battery pack modeling is  complex, as it may require consideration of electrochemistry, electrical system, thermal behavior, and control (BMS that is responsible for charge equalization, thermal management, safety etc.)  \cite{fotouhi2016review, cordoba2015control,Cai2019}. In this paper we are focused on describing the systems aspects of the battery pack, and in particular the interaction between the electrical and thermal behavior of the elements with sensing and monitoring systems. Equivalent electrical circuit models (ECMs) and lumped-parameter thermal models \cite{plett2015battery, freudiger2019generalized} are usually adopted for system level fault diagnosis, thanks to the possibility of locating voltage, current and temperature sensors. 
For simplicity, a zeroth order ECM is employed as the basis of the analysis of this paper. Note that the methodology proposed in this paper can be extended to higher order ECMs. The equations below describe a basic electrothermal model of the generic $ij$ battery cell. The model is composed of 4 equations ($e_{1},e_{2},e_{3},e_{4}$).
\begin{eqnarray}
\label{zeroECM}
e_{1}: {V_{ij}} = {V_{oc,ij}} - {R_{ij}}{I_{ij}}
\end{eqnarray}
\begin{eqnarray}
\label{SOC}
e_{2}: \frac{{dSo{C_{ij}}}}{{dt}} =  - \frac{{{I_{ij}}}}{{{Q_{ij}}}}
\end{eqnarray}
\begin{eqnarray}
\label{OCV}
e_{3}: {V_{oc,ij}} = f(So{C_{ij}})
\end{eqnarray}
\begin{eqnarray}
\label{T}
e_{4}: m{c_p}\frac{{d{T_{ij}}}}{{dt}} = {R_{ij}}{\left( {{I_{ij}}} \right)^2} - {Q_{TM{S_{ij}}}}
\end{eqnarray}
where, $V$ represents the cell terminal voltage, $I$ represents the input current. Eq.(\ref{zeroECM}) describe the zeroth order ECM. Eq.(\ref{SOC}) is the calculation of state of charge ($SoC$) using the Coulomb counting method where, $Q$ is the cell capacity. The open circuit voltage ($V_{oc}$) is a function of the $SoC$, as shown in Eq.(\ref{OCV}). Based conservation of energy, Eq.(\ref{T}) correlates the temperature $T$ to the heat generation (${R}{I}^2$) and the heat extracted by the thermal management (${Q_{TMS}}$). $m$ in Eq.(\ref{T}) represents cell mass and $c_{p}$ the specific heat capacity at constant pressure. For the purpose of structural analysis, we assume that the $R$ and $Q$ are constant and not dependent on $SoC$ and $T$. 

For the battery pack architectures shown in Fig. \ref{twotopologies}, an electrical model can be derived applying Kirchhoff's Laws and including a load current $I_{BP}$ \cite{Cai2019}.
\par For $m$P$n$S topology:
\begin{eqnarray}
\label{psKCL}
\sum\limits_{j = 1}^m {{I_{ij}} = I_{{i}}^M = {I_{BP}}\quad \quad} (i = 1 \ldots n)
\end{eqnarray}
\begin{eqnarray}
\label{psKVL}
{V_{i1}} =  \cdots  = {V_{ij}} =  \cdots  = {V_{im}} = V_{{i}}^M\quad(\forall i = 1 \ldots n )
\end{eqnarray}
where, $I_{i}^M$ and $V_{i}^M$ represent the current and voltage of the $ith$ module. $n$ indicates the number of elements in series and $m$ the number of elements in parallel.\par
For $n$S$m$P topology:
\begin{eqnarray}
\label{spKCL}
\sum\limits_{j = 1}^m {I_{{j}}^M = {I_{BP}}} \quad ({I_{ij}} = {I_{{j}}^M}\quad \forall i = 1 \ldots n)
\end{eqnarray}

\begin{eqnarray}
\begin{array}{l}
\label{spKVL}
\sum\limits_{i = 1}^n {{V_{i1}} =  \cdots }  = \sum\limits_{i = 1}^n {{V_{ij}} =  \cdots }  = \sum\limits_{i = 1}^n {{V_{im}}} = V_{{j}}^M = V_{BP} \quad  \\(\forall j = 1 \ldots m)
\end{array}
\end{eqnarray}

where, $I_{j}^M$ and $V_{j}^M$ represent the current and voltage of the $jth$ module.\par
\subsection{Structural model of battery}

\begin{figure}[]
\centering
\includegraphics[width=5cm]{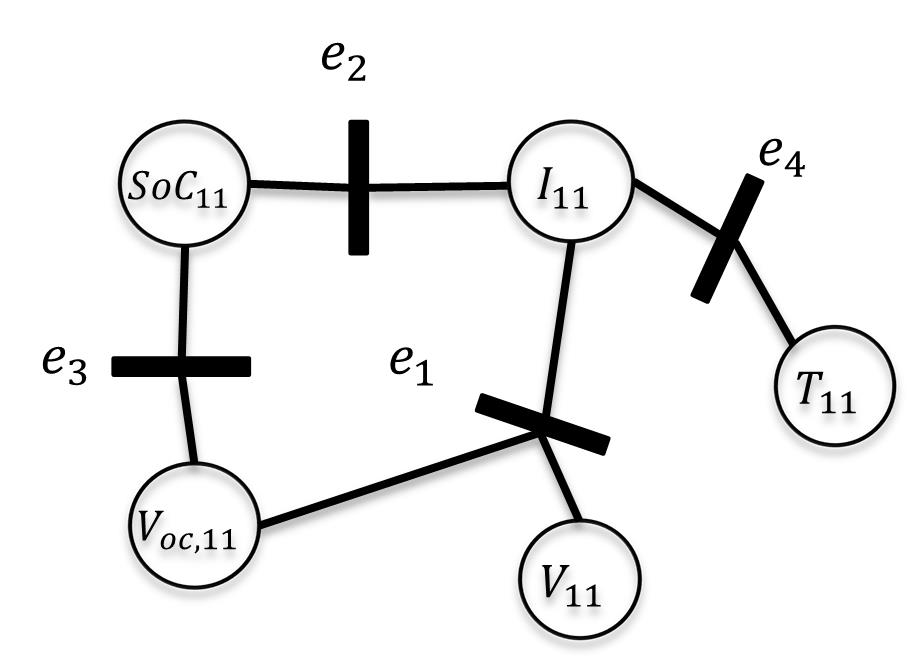}
\caption{Bipartite graph for a single cell system.}
\label{Bipartitegraph for a single cell}
\end{figure}  

\begin{table}[t!]
\caption{Incidence matrix for a single cell system}
\label{T1}
\centering
\begin{tabular}{cccccc}
\toprule
\textit{Equations}	&  \multicolumn{5}{c}{\textit{Unknown variables}} \\
\midrule
		     & $V_{11}$   & $V_{oc,11}$   & $I_{11}$    & $SoC_{11}$    & $T_{11}$\\
$e_1$		 & 1    & 1    & 1    & 0    & 0\\
$e_2$		 & 0    & 0    & 1    & 1    & 0\\
$e_3$		 & 0    & 1    & 0    & 1    & 0\\
$e_4$		 & 0    & 0    & 1    & 0    & 1\\
\bottomrule
\end{tabular}
\end{table}
Structural analysis investigates the model constraint structure \cite{Blanke2006}, i.e., the connections between known variables, unknown variables, and faults.  No matter what type of model is used, one can generate a corresponding structural model in the form of a bipartite graph. These mathematical equations can be a set of algebraic or differential equations that describe the relationships among variables. A structural representation is a bipartite graph with a set of system equations, variables and edges ($E$, $Z$, and $\varepsilon$, respectively). The set of variables ($Z$) include unknown variables ($X$) and known variables ($K$). 
\par The equations for a single cell are listed in Eq.s \eqref{zeroECM}-\eqref{T}, where $i=1,j=1$. The model has 4 equations $E =$\{$e_1, e_2, e_3, e_4$\} and 5 unknown variables $X =$\{$V_{11}, I_{11}, V_{oc,11}, SoC_{11}, T_{11}$\}. The bipartite graph of the single cell system is shown in Fig. \ref{Bipartitegraph for a single cell}, where variables are represented by circles while the equations  are represented by bars. An edge connects a variable and an equation and it is not oriented. A structural model may also be represented by a corresponding incidence matrix in which the rows represent system equations and the columns represent variables.  The elements of the incidence matrix are defined as follows: if a variable appears in an equation, the element is $1$, otherwise $0$. The incidence matrix of the single cell system is shown in Table \ref{T1}.

\subsection{mPnS versus nSmP }

\begin{figure}[htb]
\centering
\includegraphics[width=7 cm]{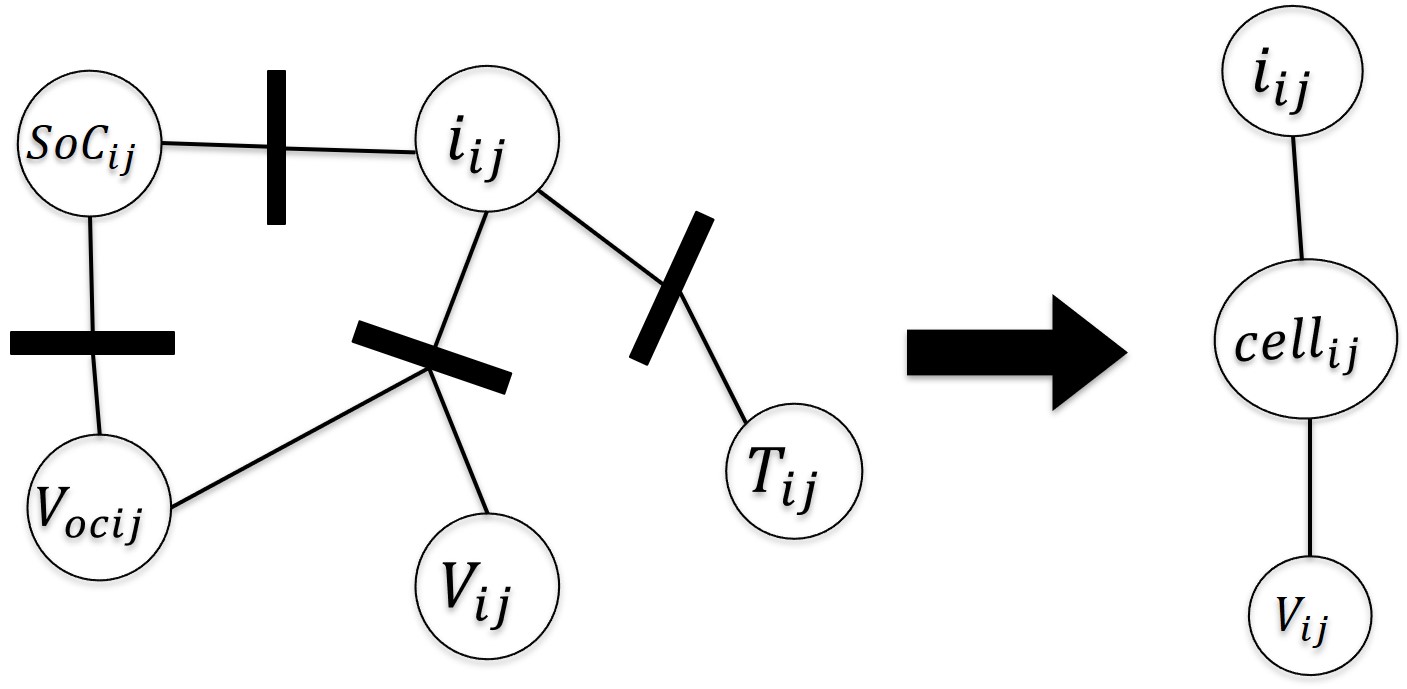}
\caption{Simplification of structural model for one cell}
\label{Simplification}
\end{figure}   

\begin{figure}[htb]
\centering
\includegraphics[width=8.5 cm]{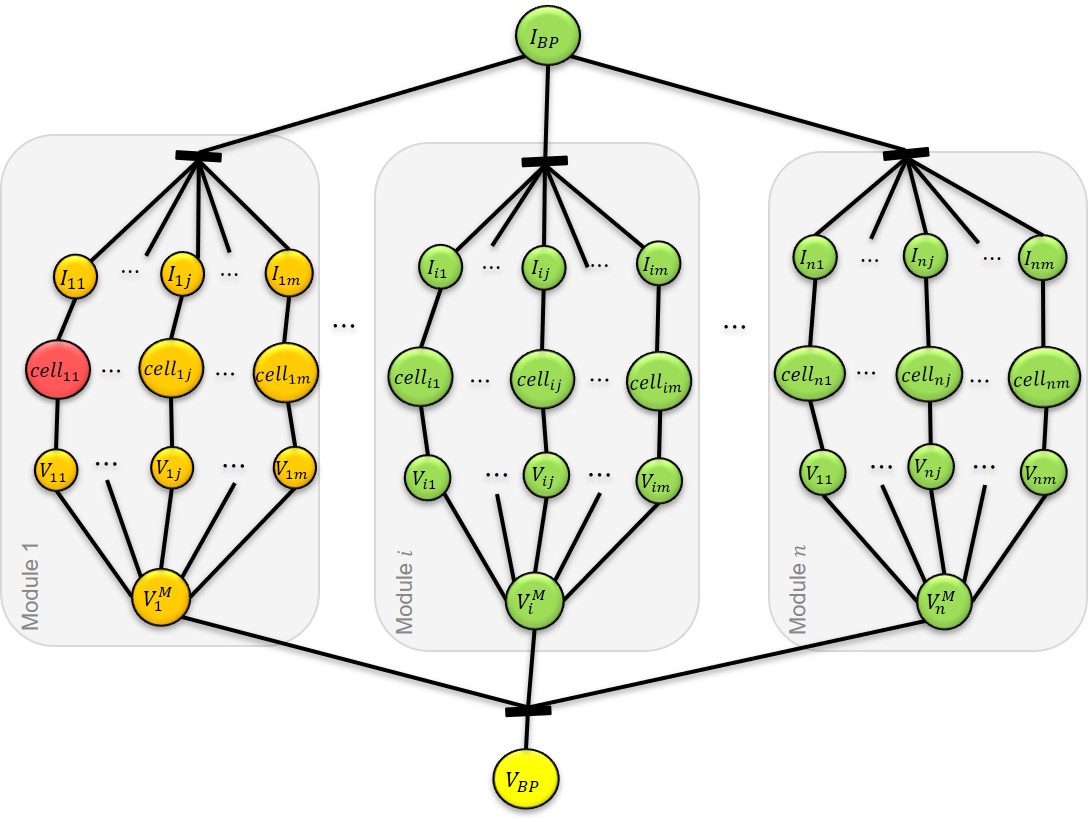}
\caption{Structural model for $m$P$n$S topology battery pack}
\label{Structural model for PS}
\centering
\includegraphics[width=8.5 cm]{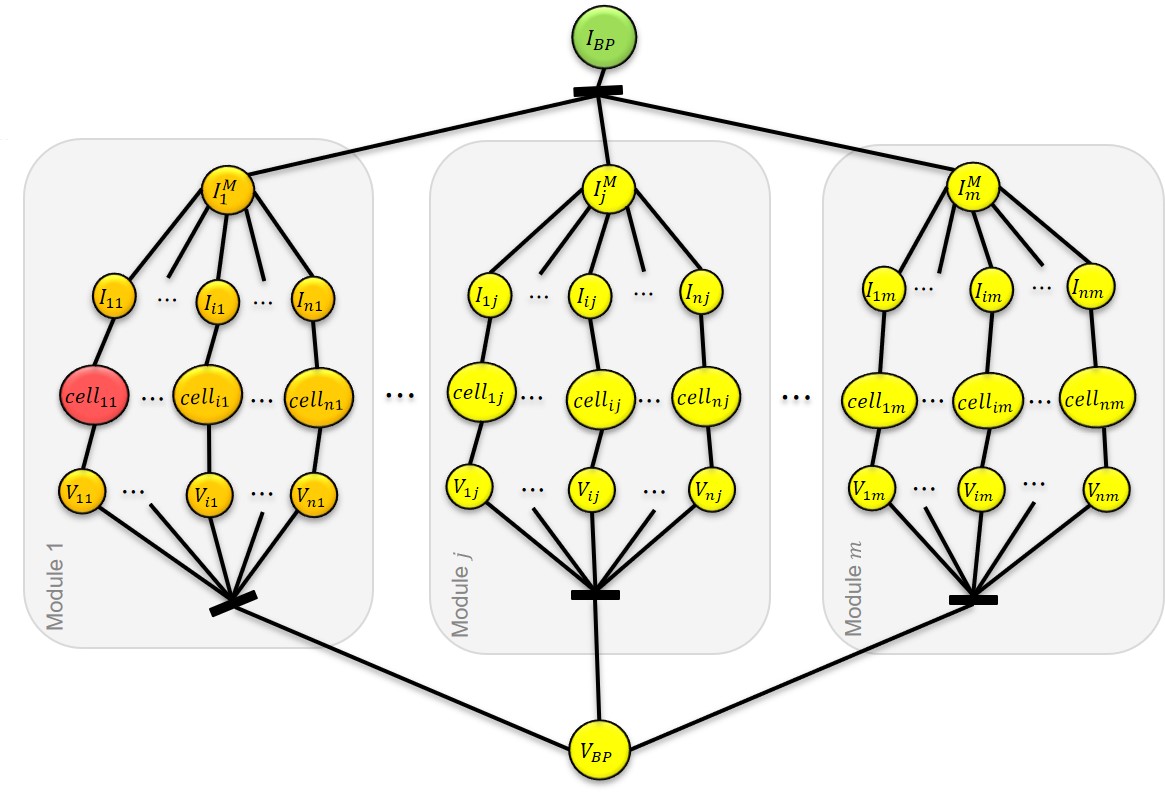}
\caption{Structural model for $n$S$m$P topology battery pack}
\label{Structural model for SP}
\end{figure} 

The $m$P$n$S and $n$S$m$P topologies are shown in Fig. \ref{twotopologies}(a) and Fig. \ref{twotopologies}(b), respectively. The system equations are listed in Eq.\eqref{zeroECM}-\eqref{spKVL}. 
Based on the battery pack model, the calculation of $SoC$, $V_{oc}$ and $T$ for one cell is isolated from another cell. There are only current and voltage connections between cells, and it is reasonable to condense the structural graph of each single cell to one node, as shown in Fig. \ref{Simplification}. With the simplified graph structure, it is possible to obtain a generalized structural model of the $m$P$n$S and $n$S$m$P topologies as shown in Fig. \ref{Structural model for PS} and \ref{Structural model for SP}, respectively, where the color coding represents the entity of the deviation between the variable of the cells. These generalized structural models can help analyze the effect of a faulty cell (for example a short circuit), or the effect of cell-to-cell variation (for example due to uneven aging of cells). Suppose that $cell_{11}$ in \textit{Module} 1 has an anomalous behavior compared to the other cells (again, due to a fault or to a change in some physical parameter) and a fixed load current is considering. For the $m$P$n$S topology, all module currents equal the pack current (see Eq. \ref{psKCL}) and remain unchanged. As shown in Fig. \ref{Structural model for PS}, any defect in $cell_{11}$ will result in an imbalance between $I_{11}, \cdots,I_{1j},\cdots,I_{1m},$ in \textit{Module} 1. The impact of the defective cell is limited to the module it belongs to and the other modules will not be affected by the defective cell. 
In the $n$S$m$P topology, the current in each module is equal to the individual cell current, and the pack current is the summation of all the module currents (see Eq. \ref{spKCL}). The variation of $cell_{11}$ will affect  $I_{11}$ and $I_{\rm{1}}^M$. Then the change of $I_{\rm{1}}^M$ will cause unbalance between $I_{\rm{1}}^M,\cdots,I_{j}^M,\cdots,I_{\rm{m}}^M$. The influence of a defective cell will therefore spread to the whole battery pack, as shown in Fig. \ref{Structural model for SP}. This is a very important intrinsic property of the two battery pack topologies, and it motivates the analysis and models presented in the following section. Note that, we only apply the simplification shown in Fig. \ref{Simplification} to the bipartite graphs of $m$P$n$S and $n$S$m$P topologies. The sensor placement analysis in the following section is based on using the system  incidence matrix without any simplification.

\section{Structural analysis for fault diagnosis}
\label{sec structual analysis}

\begin{figure}[]
\centering
\includegraphics[width=8.5cm]{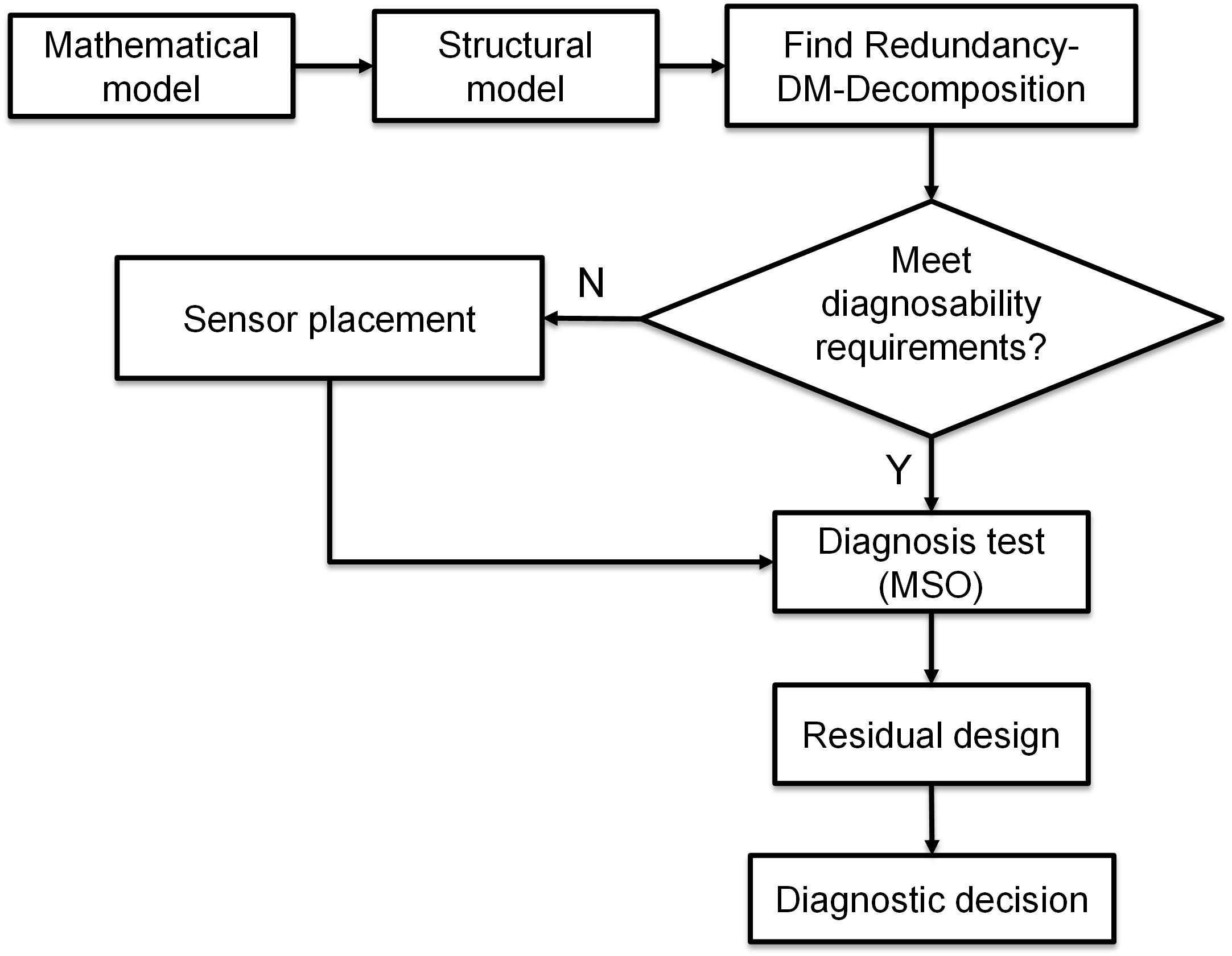}
\caption{Structural methodology for fault diagnosis}
\label{Structural methodology}
\end{figure}
\par
\begin{figure}[htb]
\centering
\includegraphics[width=5 cm]{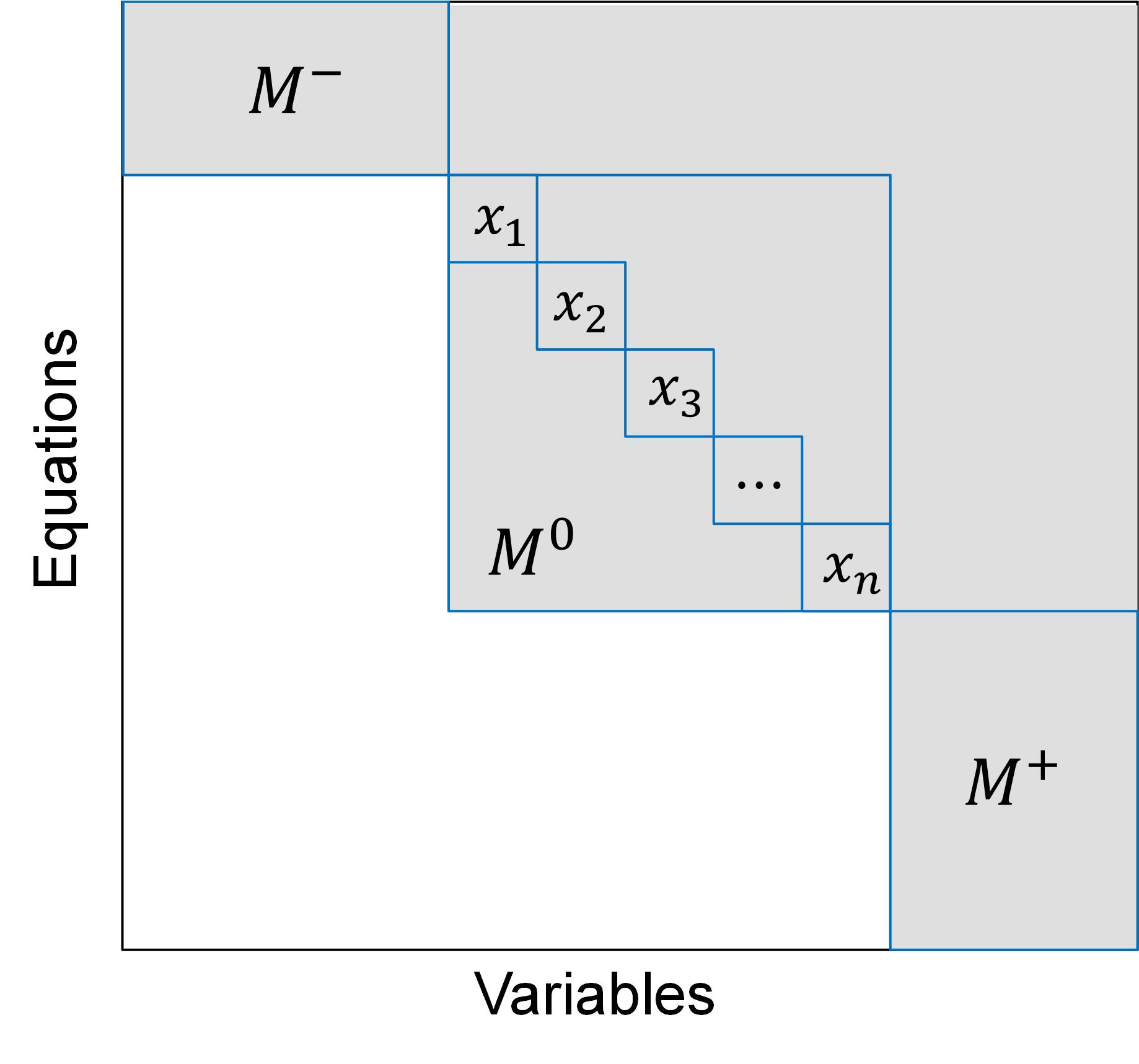}
\caption{Dulmage-Mendelsohn decomposition of a model\cite{krysander2007efficient}}
\label{DM}
\end{figure}

 The process of  structural analysis for fault diagnosis is illustrated in Fig. \ref{Structural methodology}. Generating the structural model of a system based on its mathematical model is the first step. Then, a Dulmage-Mendelsohn (DM) decomposition rearranges the system incidence matrix and divides the structural model into three subsystems \cite{dulmage1958coverings} (as shown in Fig.\ref{DM}):
\begin{enumerate}
\item under-determined part ($M^-$), where there are fewer equations than unknown variables; \item just-determined part ($M^0$), where there are equal equations as unknown variables;  \item over-determined part ($M^+$), where there are more equations than unknown variables. 
\end{enumerate}
Based on the DM decoposition of the system incidence matrix,  fault detectability and isolability are stated as follows \cite{krysander2008sensor}:
 \par \textit{fault detectability} : A fault is said to be structurally detectable if the equation containing the fault exists in the over-determined portion of the structural model decomposition: ${e_{f}} \in {\left\{ {M} \right\}^ + }$.
 \par \textit{fault isolability} : A fault $f_{i}$ is said to be structurally isolable from fault $f_{j}$, if the equation containing  $f_{i}$ exists in the over-determined part of the structural model excluding the equation that $f_{j}$ is contained: ${f_i} \in {\left\{ {M\backslash {e_{{f_j}}}} \right\}^ + }$. In other words, the equations containing these two faults must be in different equivalence classes of the over-determined subsystem. A more detailed explanations of equivalence classes may be found in \cite{krysander2007efficient}.
 \par 
If the system structural model has a sufficient degree of redundancy to meet the requirements of FDI, different proper Minimal Structurally Over-determined (MSO) subsystems can be found and used to design diagnostic tests, that is, algorithms that are capable of verifying the consistency of the equations. Any inconsistency could be attributed to the presence of a fault, but could of course also be due to noise and uncertainty. A diagnosis test is an over-determined subsystem of the complete system model where one (or more) fault(s) is (are) structurally detectable \cite{krysander2007efficient}. A MSO set is a proper over-determined set of model equations such that its degree of redundancy is one, i.e: $\left|E\right|-\left|X\right|=1$. Within each MSO set, by finding proper matching, a residual can be found that is only sensitive to a certain subset of faults. A residual is a signal used as a fault indicator that is sensitive to these faults. Mapping the residuals to each corresponding fault results in a fault signature matrix; solutions can then be found to make every fault detectable or uniquely isolable from other faults as required.
\par If the system structural model has no over-determined subsystem, then it is not possible to detect or isolate any faults. Utilizing the definitions of detection and isolation, we can create a sensor installation guide that allows us to generate the desired degree of redundancy to develop FDI algorithms based on the system diagnostic specifications.

\subsection{Analytical redundancy (AR) in a battery system}
\label{AR}
When no measurement is considered, a single cell system of Eq.s \eqref{zeroECM}-\eqref{T} has five unknown variables ($X=5$) and four equations ($E=4$). The system is under-determined and  cannot be solved to calculate the five unknown variables. Thus, the intrinsic analytical redundancy ($IAR$) of a single cell is $-1$.
\begin{eqnarray}
\label{Lemma1}
IAR_{single\,cell}=-1
\end{eqnarray}
In the same way, the $IAR$ of the battery system is also $-1$ for both $m$P$n$S and $n$S$m$P topologies, when no sensors and faults are considered because the number of equations is always 1 less than the number of unknown variables.
\begin{eqnarray}
\label{Lemma2}
IAR_{battery\, pack}=-1
\end{eqnarray}
To increase the degree of AR and provide the ability to design diagnostic algorithms, sensors are needed in the battery system. In the following subsections, we use graph-theoretic tools to understand how different measurements (current, voltage, and temperature) can add analytical redundancy to the system, and how this analytical redundancy is linked to system diagnosability.
\par The addition of a sensor to measure an unknown variable increases the number of known variables and equations, however it introduces the possibility of a fault, as shown in the following equation:
\begin{eqnarray}
\label{sensorEQ}
{y_{u}}= u + {f_{y_{u}}}
\end{eqnarray}
where $y$ denotes the real time sensor reading and is a known variable, $f_{y_{u}}$ denotes the sensor fault ($f \ne 0$  indicates that the sensor failed). $u$ is the actual value of the sensed current, voltage or temperature. 

\subsection{Matching on a structural model}
\begin{figure}[]
\centering
\includegraphics[width=8.5cm]{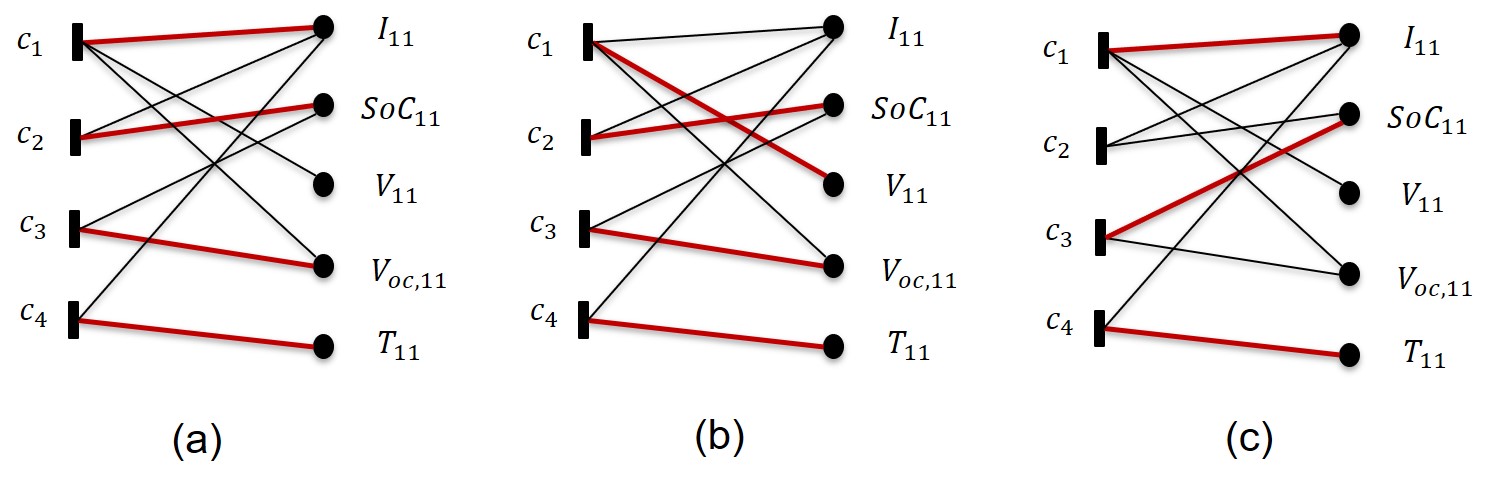}
\caption{Two complete matchings (a) , (b) and an incomplete matching (c) }
\label{Three_matchings}
\end{figure} 
The basic principle of structural analysis is to find \textit{matchings}, that is, causal assignments between $X$ and $E$ in a structural model. If an $X$ is matched with a $E$, it can be calculated from this equation. If an $X$ is not matched, it cannot be calculated. A more detailed definition of matching can be found in \cite{Blanke2006}. Basically, if we employ bipartite graph as the structural model, a matching is a subset of $\varepsilon$. Any two edges in a matching do not share common node ($E$ or $Z$), which means it associates one equation ($E$) with one specific variable ($Z$). 
Matching is not unique, as different matchings may be found for a system. Fig. \ref{Three_matchings} lists three possible matchings for the single cell system Eq.s \eqref{zeroECM}-\eqref{T}. The black thinner lines represent unmatched edges, while the red bold lines represent matched edges. A matching can be further defined as a complete matching based on the number of edges ($\left|\varepsilon\right|$), equations ($\left|E\right|$), and variables ($\left|Z\right|$) contained in the matching. A matching is said to be : i) complete with respect to $E$ if $\left|\varepsilon\right|=\left|E\right|$; ii) complete with respect to $Z$ if $\left|\varepsilon\right|=\left|Z\right|$; iii) if only unknown variables ($X$) are considered, a \textit{matching} is said to be complete if $\left|\varepsilon\right|=\left|X\right|$. In Fig. \ref{Three_matchings}, (a) and (b) are two complete matchings with respect to equations; (c) is an incomplete matching. 
When no measurement is considered, a single cell system has less equations than unknown variables, thus a complete matching can be find only with respect to equations.

\begin{figure}[t!]
\centering
\includegraphics[width=8.5cm]{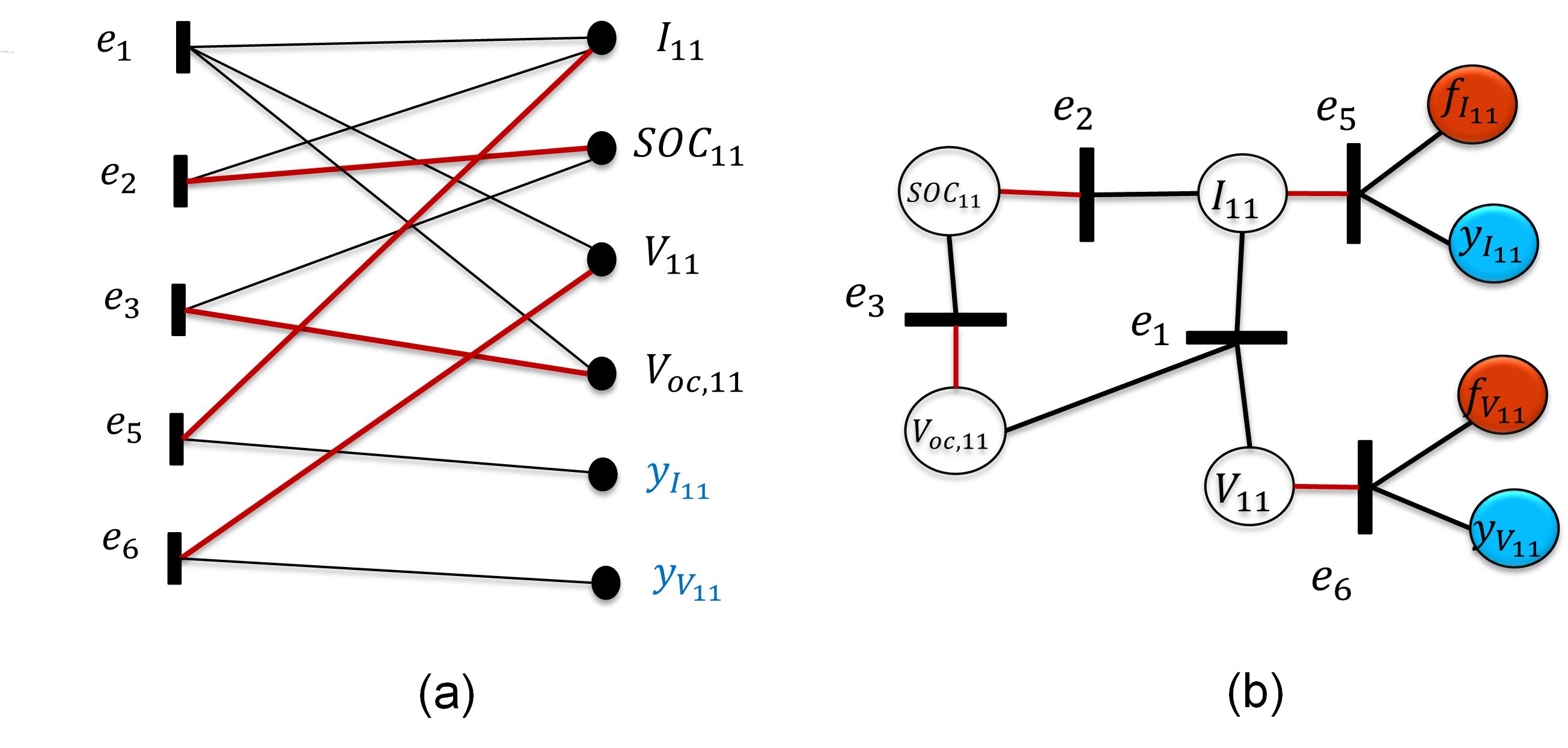}
\caption{  (a) an example of complete matching, (b) matching in the bipartite graph }
\label{ matching in the bipartite graph}
\end{figure}

As example, a complete matching of the system of Eq.s \eqref{zeroECM}-\eqref{T} with respect to $X$ can be achieved by adding the following sensor equations to measure current and voltage of the single cell:
\begin{eqnarray}
\label{current sensor}
{e_5}:{y_{I_{11}}} = I_{11} + {f_{y_{I_{11}}}}
\end{eqnarray} 
\begin{eqnarray}
\label{voltage sensor}
{e_6}:{y_{V_{11}}} = {V_{11}} + {f_{y_{V_{11}}}}
\end{eqnarray} 
where $y_{I_{11}}$ and $y_{V_{11}}$ denotes the real time current and voltage sensor reading and are known variables, $f_{y_{I_{11}}}$ and $f_{y_{V_{11}}}$ denotes the sensor faults. The single cell model with a current and a voltage measurement includes 6 equations  $E =$\{$e_1, e_2, e_3, e_4, e_5, e_6$\} and 5 unknown variables $X =$\{$V_{11}, I_{11}, V_{oc,11}, SoC_{11}, T_{11}$\}. There are more equations than unknown variables. The degree of AR in this case is 1, indicating that the single cell system with a current sensor and a voltage sensor has an over-determined subset of equations and it is therefore possible to design a diagnostic test. 
\begin{eqnarray}
\label{Lemma3}
AR_{single \, cell}=1 \quad when \; \exists \; y_{I_{11}} \; and \; y_{V_{11}}
\end{eqnarray}
A MSO set $MSO =$\{$e_1, e_2, e_3, e_5, e_6$\} can be found from the 6 equations. 
A complete matching with respect to unknown variables can be found in this MSO set, as shown in Fig. \ref{ matching in the bipartite graph}(a). The bipartite graph of  Fig. \ref{ matching in the bipartite graph}(b) shows this matching. The red lines represent the matched edges, blue circles represent known variables and red circles represent faults.

\subsection{Oriented graph}

\begin{figure}[t!]
\centering
\includegraphics[width=6cm]{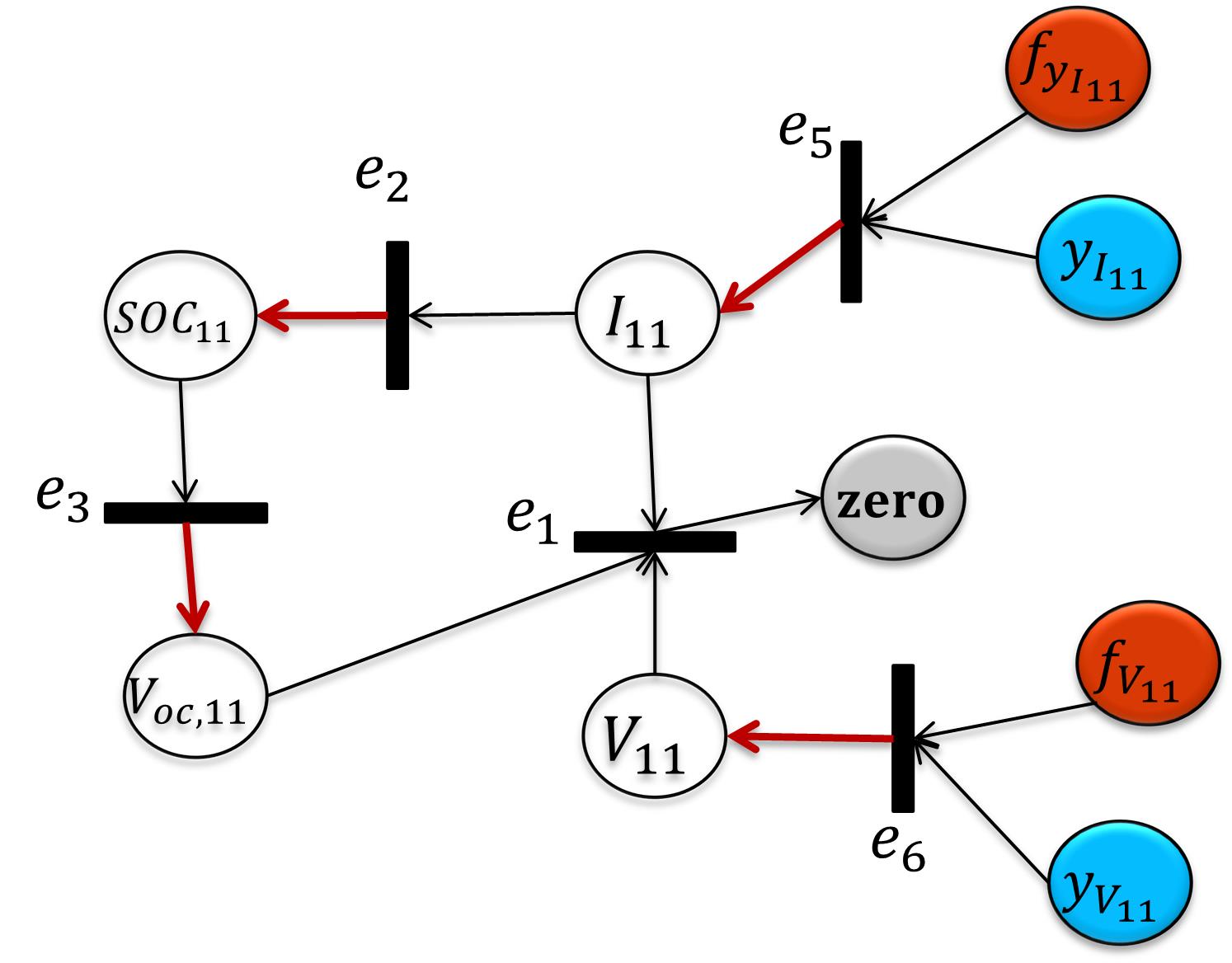}
\caption{Oriented graph for single cell system with a current and a voltage measurement }
\label{Oriented_graph}
\end{figure}

An oriented graph is a matching that assigns orientation of some edges. For a matched equation, the edge that connects the matched variable and the equation is called a matched edge directed from the equation to the variable. Other edges that connect non-matched variables and  equations are called non-matched edges with an orientation from non-matched variables to the equation. For equations that are not matched, all edges' orientations are from variables to the equation. The non-matched equation generates a zero output, which represents an analytic redundant relation (ARR) of the model. ARRs are used to generate residuals as fault indicators for the purpose of fault diagnosis. Fig. \ref{Oriented_graph} shows the oriented structural graph for the MSO set found in a single cell system with a current and a voltage measurement. 

The oriented graph defines a set of computational sequences $S=\{S_{1},S_{2}\}$ to calculate the unknown variables:
  
  $S_{1}=\left\{\left(e_5,\
  I_{11}\right),\ \left(e_2,\ SoC_{11}\right),\ \left(e_3,\ V_{oc,11}\right) \right\}$
  
  $S_{2}=\left\{\left(e_6,\
  V_{11}\right) \right\}$
  
\noindent where, the pair $(e,\ x)$ means variable $x$ is computed from equation $e$. The order of the pairs defines a computational sequence. Note that $e_{2}$ is a differential equation, and when we use their integral causalities, the knowledge of initial values are required. The oriented graph results in an alternated chain that starts from the known variables and alternates successively between two nodes \cite{Blanke2006}. For the oriented graph shown in Fig. \ref{Oriented_graph}, the alternated chain based on the computational sequence $S_{1}$ can be expressed as: 
\begin{eqnarray}
    \label{alternated chian}
  y_{I_{11}}\rightarrow e_5\rightarrow I_{11}\rightarrow e_2\rightarrow SOC_{11}\rightarrow e_3\rightarrow V_{oc,11}
\end{eqnarray} 

Based on the alternated chain, the structural reachability is defined as \cite{Blanke2006}: a variable $z_{2}$ is reachable from a variable  $z_{1}$ if there exists an alternated chain from  $z_{1}$ to  $z_{2}$ . The circle in gray with a `zero' in it represents the ARR of the model. In Fig. \ref{Oriented_graph}, $e_{1}$ is the ARR in this MSO with the matching we choose in Fig. \ref{ matching in the bipartite graph}. A residual based on the sensor set $\{y_{I_{11}}, y_{V_{11}}\}$ that is capable of detecting the two faults $\{f_{y_{I_{11}}}, f_{y_{V_{11}}}\}$ is

\begin{eqnarray}
\begin{array}{l}
\label{residual}
r=y_{V_{11}}-f[So{C_{11,0}} - \frac{1}{Q}\int_0^t {y_{I_{11}}(t)dt}]+Ry_{I_{11}}\\=f_{y_{V_{11}}}+Rf_{y_{I_{11}}}
\end{array}
\end{eqnarray}

\noindent where, $So{C_{11,0}}$ represents the initial value of $So{C_{11}}$. The residual $r$ in Eq.\eqref{residual} is obtained by substituting all matched equations to $c_{1}$ to eliminate unknown variables and make it only contains known variables. A violation of any equation that is used to generate the residual will result in a non-zero residual indicating a fault. In fact, the residual in Eq.\eqref{residual} is the only residual generator for a single cell with current and voltage measurements for the matching of Fig. 9. When there isn't a fault, $r$ should be $0$. Notice that this residual is sensitive to both $f_{y_{I_{11}}}$ and $f_{y_{V_{11}}}$.

\subsection{Oriented graph of a single cell with different measurements}
\label{subsec singlecell diffsen}

\begin{figure}[t]
\centering
\includegraphics[width=8.5cm]{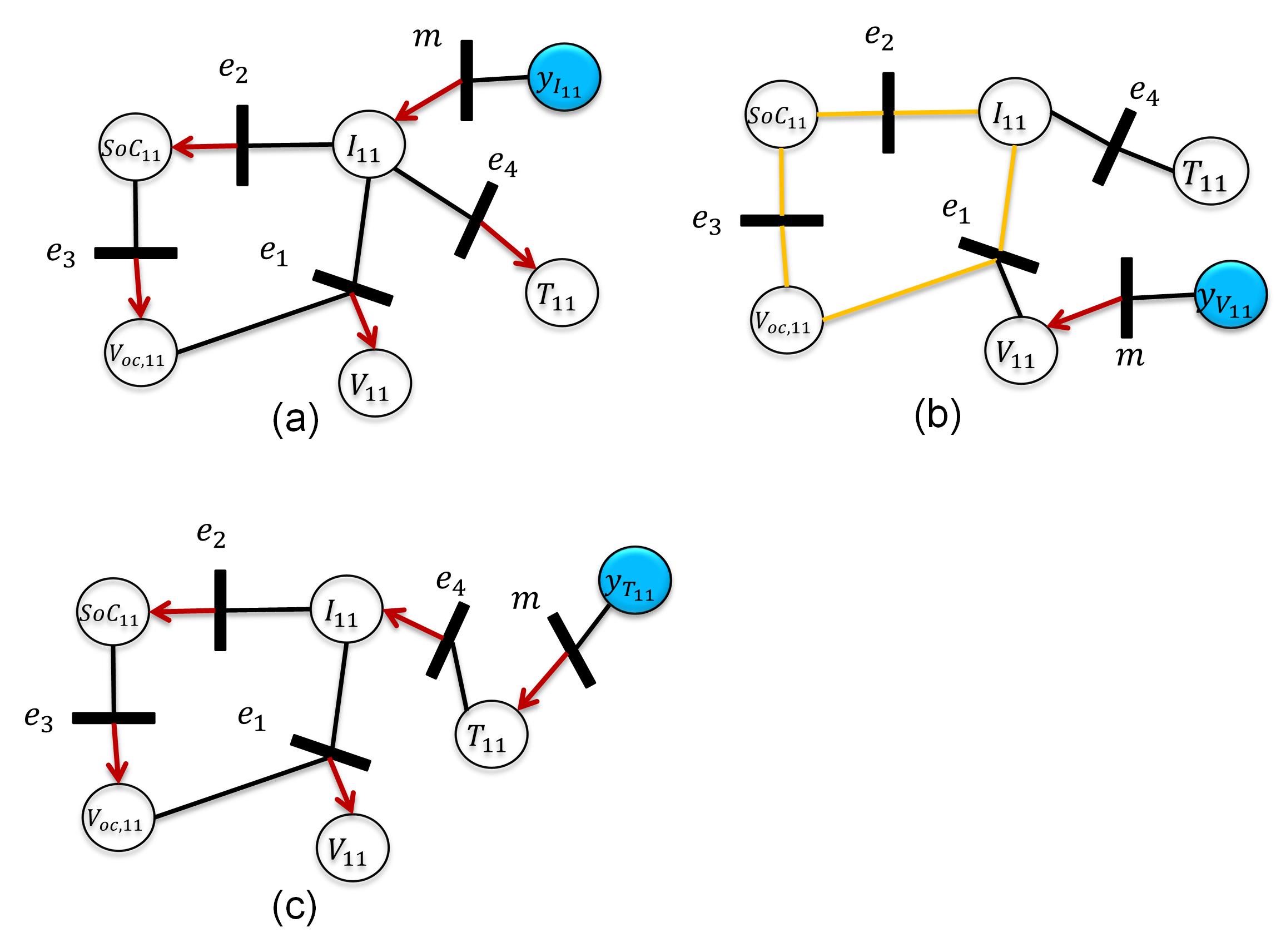}
\caption{Oriented structural graph of a single cell (a) with one current measurement   (b) with one  voltage measurement, (c) with one temperature measurement}
\label{different sensing}
\end{figure}
The structural model of a single cell represented by a bipartite graph is shown in Fig. \ref{Bipartitegraph for a single cell}. Without sensing, we cannot solve for the unknown variables. If we have a current measurement for a single cell, the resulting structural graph is shown in Fig. \ref{different sensing}(a). Every unknown variable is easily reachable from a measurement (known) because an alternated chain can be found to exist and this is always true when higher order ECM is employed.  If we introduce a voltage measurement for a single cell, then the structural graph is as shown in Fig. \ref{different sensing}(b). Notice that in this case, the three equations $\left\{e_1,e_2,e_3\right\}$ form a loop which requires the three equations to be solved simultaneously and this is always true when higher order ECM is employed. If we have a temperature measurement for a single cell, the resulting structural graph is shown in Fig. \ref{different sensing}(c). There is an alternated chain to make every unknown variable reachable from the measurement but this is true when we use zeroth order ECM. If a higher order ECM is employed to represent the battery, adding a temperature measurement will result in a loop as in the case of adding a voltage measurement. This can be proved by listing the model equations using higher order ECM and drawing its corresponding bipartite graph. Due to the limitation of paper length, the proof is not presented here. While it is true that in the case of a voltage or temperature measurement, the redundancy of single cell system is increased by 1 (becomes 0 from $-1$) as well and we can still calculate all unknown variables, it is not as easy to compute these variables as was the case with a current measurement.
This indicates that, in principle, current sensors allow us to implement residual generators with lower computational burden compared to voltage and temperature sensors. This property is useful to select a sensor at the cell level in the next section, when we develop sensor placement strategies for a battery pack to achieve FDI.


\section{Sensor placement for fault detectability and isolability analysis}
\label{sec sensor placement}

In this section, we develop a systematic methodology to find the minimal sensor sets that can potentially provide the two common battery topologies with enough ARRs to develop diagnostic algorithms that can achieve complete fault isolation. In other words, with the sensor installation guide developed in this section, it is possible to design algorithms to find MSO sets and generate residuals, each sensitive to a unique fault. \par 
As discussed in Section \ref{AR}, if sensors are not included in a battery system (whether consisting of a single cell or of $m$P$n$S and $n$S$m$P topologies), the system is under-determined (Eqs. \eqref{Lemma1} and \eqref{Lemma3}), and there is no analytical redundancy in it to permit diagnosis. It is clear, then, that sensors are necessary to achieve analytical redundancy in a battery pack. As example, by adding two sensors the system will have an over-determined subsystem. The addition of different sensor types in different locations will result in the generation of different over-determined subsystems. Different combinations of sensors and the presence of different faults will give rise to different \textit{fault detectability} properties, and it may take more than two sensors to insure detectability of all faults. If faults are detectable by adding the appropriate sensors, it will then possible to generate a residual. Based on the residual, the \textit{fault isolability} can be defined in another way \cite{jung2020sensor}: fault $f_{i}$ is isolable from fault $f_{j}$, if there exists a residual that is sensitive to $f_{i}$ but not $f_{j}$. Indeed from Eq.\eqref{residual}, it can be found that in a single cell instrumented with a current sensor and a voltage sensor, the two sensor faults are detectable but are not isolable from each other.  
\par Detectability and isolability analysis can be easily performed using the Structural Analysis Toolbox developed by Frisk et al. \cite{frisk2017toolbox}. In the next subsection, the faults considered in this study are introduced and a fault detectability and isolability analysis is performed for a single cell, the generalized $m$P$n$S and $n$S$m$P topologies.

\subsection{Battery modeling with faults}

A battery pack can exhibit anomalous behavior due to many reasons, including short circuit (internal or external to the cell), resistance increase and/or capacity fade due to accelerated aging, sensor fault, or BMS fault \cite{bubbico2018hazardous}. In this work, two types of faults are considered: sensor fault and short circuit faults; these may occur at the cell or module level in the battery pack. Short circuit faults are especially important because, unlike other anomalies that would still permit the system to operate (e.g. battery degradation), a short circuit may lead to thermal runaway and result in a catastrophic failure.

\begin{figure}[t!]
\centering
\includegraphics[width=5 cm]{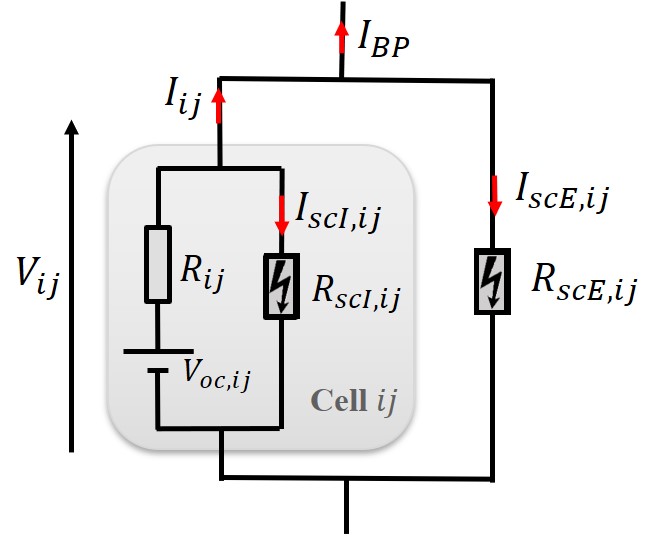}
\caption{Diagram of internal and external short circuit in a cell}
\label{Diagram of SC}
\end{figure}

Internal and external short circuit faults are depicted in the circuit diagram of Fig. \ref{Diagram of SC}. The internal short circuit is represented by a parallel resistance ($R_{scI}$) connected to the cell \cite{ouyang2015internal}. The external short circuit is similarly represented by a parallel resistance ($R_{scE}$) externally connected to a cell or a module. The fault model for internal short circuit is given by:
\begin{eqnarray}
\label{ISC}
{I_{scI,ij}} = \left( {\frac{V_{ij}}{{{R_{scI}}}}} \right){f_{scI,ij}}
\end{eqnarray}
where, $f_{scI,ij}$ represents the internal short circuit current fault and is a binary variable with value 1 when the fault is present, and  $I_{scI,ij}$ the internal short circuit current. The model for a fault free battery is depicted by Eq.s \eqref{zeroECM}-\eqref{T}. In the case of an internal short circuit, Eq. \eqref{OCV} remains the same while Eq.s \eqref{zeroECM}, \eqref{SOC},and \eqref{T} result in the following equations \eqref{zeroECMf}-\eqref{Tf}.

\begin{eqnarray}
\label{zeroECMf}
{V_{ij}} = {V_{oc,ij}} - {R_{ij}}({I_{ij}+I_{scI,ij}})
\end{eqnarray}
\begin{eqnarray}
\label{SOCf}
\frac{{dSo{C_{ij}}}}{{dt}} =  - \frac{{({I_{ij}+I_{scI,ij}})}}{{{Q_{ij}}}}
\end{eqnarray}
\begin{eqnarray}
\label{Tf}
m{c_p}\frac{{d{T_{ij}}}}{{dt}} = {R_{ij}}{\left( {{I_{ij}+I_{scI,ij}}} \right)^2} - {Q_{TM{S_{ij}}}}
\end{eqnarray}

The fault model for the external short circuit is:
\begin{eqnarray}
\label{ESC}
{I_{scE}} = \left( {\frac{V_{E}}{{{R_{scE}}}}} \right){f_{scE}}
\end{eqnarray}

\noindent where, $f_{scE}$ (a binary variable) represent the external short circuit fault, and $I_{scE}$ the external short circuit current. When we consider a module, that is the composition of multiple cells, then the voltage $V_{E}$ across the short circuit resistance $R_{scE}$  depends on how many cells are shorted by the external short circuit. For example, as shown in Fig. \ref{SP diagram with faults} and \ref{PS diagram with faults} later, if we consider the external short circuit at the module level, for the external short circuit in \textit{Module} $i$ in $m$P$n$S topology,  $V_{E,i}=V_{i1}=\cdots=V_{ij}=\cdots=V_{im}$; for the external short circuit in \textit{Module} $j$ in $n$S$m$P topology,  $V_{E,j}=V_{1j}+\cdots+V_{ij}+\cdots+V_{nj}$.  $I_{scE}$ will appear in the KCL equations, (Eq. \eqref{spKCL} or \eqref{psKCL}). For example, for a single cell system as shown in Fig.\ref{Diagram of SC}, the external short circuit fault can be modeled as:
\begin{eqnarray}
\label{ESC11}
{I_{scE,11}} = \left( {\frac{V_{11}}{{{R_{scE}}}}} \right){f_{scE,11}}
\end{eqnarray}
\begin{eqnarray}
\label{KCL11}
{I_{11}} = {I_{BP}} + {I_{scE,11}}
\end{eqnarray}

\noindent The redundancy of the battery system model with short circuit faults remains $-1$, because the addition of an unknown variable to the system is balanced by the introduction of a new equation. 
\begin{eqnarray}
\label{Lemma4}
AR_{model \, with \, faults}=-1 \quad 
\end{eqnarray}
\par Sensor faults modeling was introduced in Section Section \ref{AR} see Eq.\ref{sensorEQ}.
\subsection{Fault detectabilty and isolability analysis and sensor placement for single cell}
The mathematical model of a single cell system with faults is shown in Eq.s \eqref{OCV}, \eqref{ISC}-\eqref{Tf}, \eqref{ESC11} and \eqref{KCL11} with $i=1, j=1$. The set of short circuit faults that are included in the model is \{$f_{scI,11},f_{scE,11}$\}. The set of sensor faults depends on what sensors are added to the battery system. For the single cell system, the possible sensor positions are \{$I_{BP},I_{11},V_{11},T_{11}$\}. 

\begin{table}[]
	\caption{Fault detectability and isolability matrix for a single cell without sensor and with one sensor (ND=Non Detectable; NA=Non Applicable) }
 	\centering
 	\label{t2}
 	\renewcommand{\arraystretch}{1.3}
\begin{tabular}{lllllll}
\toprule
               & $f_{scI,11}$  & $f_{scE,11}$  & $f_{y_{I_{BP}}}$  & $f_{y_{I_{11}}}$  & $f_{y_{V_{11}}}$   & $f_{y_{T_{11}}}$   \\ \midrule
no
sensor & ND & ND & NA & NA & NA & NA \\ 
\{$y_{I_{BP}}$\}              & ND & ND & ND & NA & NA & NA \\ 
\{$y_{I_{11}}$\}             & ND & ND & NA & ND & NA & NA \\ 
\{$y_{V_{11}}$\}              & ND & ND & NA & NA & ND & NA \\ 
\{$y_{T_{11}}$\}              & ND & ND & NA & NA & NA & ND \\ \bottomrule
\end{tabular}
\end{table}
\begin{figure*}[htb]
\centering
\includegraphics[width=16cm]{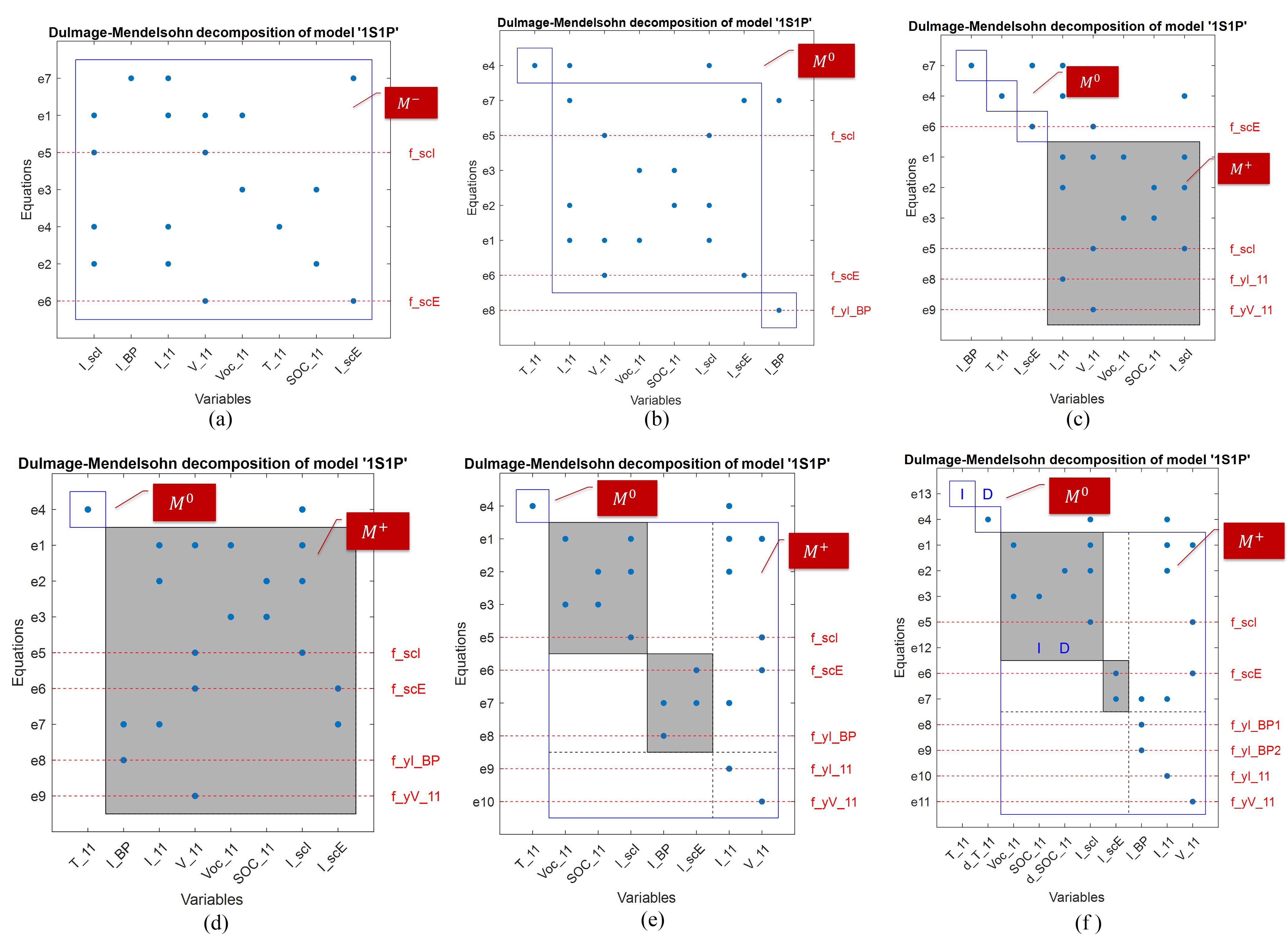}
\caption{DM decompositions of a single cell system (1S1P or 1P1S) (a) without sensor or with sensor set:  (b) \{$y_{I_{BP}}$\}, (c) \{$y_{I_{11}},y_{V_{11}}$\}, (d) \{$y_{I_{BP}},y_{V_{11}}$\}, (e) \{$y_{I_{11}},y_{V_{11}},y_{I_{BP}},$\}, (f) \{$y_{I_{11}},y_{V_{11}},y_{I_{BP1}},y_{I_{BP2}}$\}}
\label{F12}
\end{figure*}

\begin{table}[]
	\caption{Fault detectability and isolability matrix for a single cell with two sensors (ND=Not Detectable; D=Detectable; NI=Not Isolable; NA=Not Applicable; ) }
 	\centering
 	\label{t3}
 	\scriptsize
 	\renewcommand{\arraystretch}{1.3}
\begin{tabular}{llllllll}
\toprule
               & $f_{scI,11}$  & $f_{scE,11}$  & $f_{y_{I_{BP}}}$  & $f_{y_{I_{11}}}$  & $f_{y_{V_{11}}}$   & $f_{y_{T_{11}}}$   \\ 
\midrule
\{$y_{I_{11}},y_{V_{11}}$\} & D,NI & ND & NA & D,NI & D,NI & NA \\ 
\{$y_{I_{11}},y_{T_{11}}$\} & D,NI & ND & NA & D,NI & NA & D,NI \\ 
\{$y_{V_{11}},y_{T_{11}}$\} & D,NI & ND & NA & ND & D,NI & D,NI \\ 
\{$y_{I_{BP}},y_{I_{11}}$\} & D,NI & D,NI & D,NI & D,NI & NA & NA \\ 
\{$y_{I_{BP}},y_{V_{11}}$\}  & D,NI & D,NI & D,NI & NA & D,NI & NA \\ 
\{$y_{I_{BP}},y_{T_{11}}$\}  & D,NI & D,NI & D,NI & NA & NA & D,NI \\ 
\bottomrule
\end{tabular}
\end{table}
\par Table \ref{t2} shows the fault detectability and isolability matrix  for the single cell system without sensors, and with only one sensor. It can be seen that all faults cannot be detectable with a single sensor.  The DM decomposition shown in Fig. \ref{F12}(a) and (b) show that without any sensor, the system is under-determined (7 equations and 8 unknowns) and with one sensor (choose sensor \{$y_{I_{BP}}$\} as an example), the system becomes just-determined.

Table \ref{t3} shows the fault detectability and isolability matrix for the single cell system with 2 sensors. There are 6 possible sensor sets. With sensor sets \{$y_{I_{11}},y_{V_{11}}$\}, \{$y_{I_{11}},y_{T_{11}}$\}, \{$y_{V_{11}},y_{T_{11}}$\}, all the internal short circuit faults and sensor faults are detectable, while the external short circuit fault is not. As shown in the DM-decomposition result (choose sensor set \{$y_{I_{11}},y_{V_{11}}$\} as an example, shown in Fig. \ref{F12}(c)): the equation containing the external short circuit fault signal ($f_{scE,11}$) is in the just-determined part, which means $f_{scE,11}$ is not detectable. Equations containing fault signals $f_{scE,11},f_{y_{I_{11}}},f_{y_{V_{11}}}$ are in the over-determined part and in the same equivalence class (gray box). Thus, these three faults are detectable but not isolable from each other. 
Table \ref{t3}  shows that with the other three sensor sets \{$y_{I_{BP}},y_{I_{11}}$\}, \{$y_{I_{BP}},y_{V_{11}}$\}, \{$y_{I_{BP}},y_{T_{11}}$\} all faults can be detectable, but are also not isolable. The DM-decomposition results of one of these sensor set illustrates this as well (choose sensor set \{$y_{I_{BP}},y_{V_{11}}$\} as an example, shown in Fig. \ref{F12}(d)): all the equations containing fault signals are in the over-determined part, which indicates all faults can be detectable. However, all equations containing fault signals are in the same equivalence class, which indicates that these faults are not isolable from each other. 

\begin{table*}[htb]
	\caption{Fault detectability and isolability matrix for a single cell with three sensors (ND=Not Detectable; D=Detectable; NI=Not Isolable; I=Isolable; UI=Uniquely Isolable; NA=Not Applicable; ) }
 	\centering
 	\label{t4}
 	\scriptsize
 	\renewcommand{\arraystretch}{1.3}
\begin{tabular}{lllllll}
\toprule
               & $f_{scI,11}$  & $f_{scE,11}$  & $f_{y_{I_{BP}}}$  & $f_{y_{I_{11}}}$  & $f_{y_{V_{11}}}$   & $f_{y_{T_{11}}}$   \\ \midrule
\{$y_{I_{11}}, y_{V_{11}}, y_{T_{11}}$\} & D,UI & ND & NA & D,UI & D,UI & D,UI \\ \hline
\{$y_{I_{11}}, y_{V_{11}}, y_{I_{BP}}$\} & D,UI & D,I & D,I & D,UI & D,UI & NA \\ \hline
\{$y_{I_{11}}, y_{T_{11}}, y_{I_{BP}}$\} & D,UI & D,I & D,I & D,UI & NA & D,UI \\ \hline
\{$y_{V_{11}}, y_{T_{11}}, y_{I_{BP}}$\} & D,UI & D,I & D,I & NA & D,UI & D,UI \\ 
\bottomrule
\end{tabular}
\end{table*}

Table \ref{t4} shows the fault detectability and isolability matrix for the 1S1P system with 3 sensors. With sensor set \{$y_{I_{11}},y_{V_{11}},y_{T_{11}}$\}, all faults can be detectable except for the external short circuit fault. With sensor sets \{$y_{I_{11}},y_{V_{11}},y_{I_{BP}}$\},\{$y_{I_{11}},y_{T_{11}},y_{I_{BP}}$\},\{$y_{V_{11}},y_{T_{11}},y_{I_{BP}}$\}, all faults are detectable. The internal short circuit fault and all sensor faults can be uniquely isolable, while the  external short circuit fault and the load current sensor fault can be isolable from other faults but these two faults cannot be isolated from one another. From the DM-decomposition result (choose sensor set \{$y_{I_{11}},y_{V_{11}},y_{I_{BP}}$\} as an example, shown in Fig. \ref{F12}(e)), it can be seen that: all faults are located in the over-determined part. The equations containing fault signals $f_{scE,11}$ and $f_{y_{I_{BP}}}$ are in the same equivalence class, which means they are not isolable from each other but they are isolable from other faults.  The equations containing fault signals $f_{scI,11}$, $f_{y_{I_{11}}}$, $f_{y_{V_{11}}}$ are in the different equivalence classes which means these faults can be uniquely isolable.

\begin{table*}[t]
	\caption{Fault detectability and isolability matrix for a single cell with four sensors (ND=Not Detectable; D=Detectable; NI=Not Isolable; UI=Uniquely Isolable; NA=Not Applicable; ) }
 	\centering
 	\setlength{\tabcolsep}{1.65mm}{
 	\label{t5}
    \renewcommand{\arraystretch}{1.3}
\begin{tabular}{llllllll}
\toprule
               & $f_{scI,11}$  & $f_{scE,11}$  & $f_{y_{I_{BP1}}}$  & $f_{y_{I_{11}}}$  & $f_{y_{V_{11}}}$   & $f_{y_{T_{11}}}$ & $f_{y_{I_{BP2}}}$   \\ \midrule
\{$y_{I_{11}}$, $y_{V_{11}}$, $y_{T_{11}}$, $y_{I_{BP}}$\} & D,UI & D,I & D,I & D,UI & D,UI & D,UI & NA \\
\{$y_{I_{11}}$, $y_{V_{11}}$, $y_{I_{BP1}}$, $y_{I_{BP2}}$\}   & D,UI & D,UI & D,UI & D,UI & D,UI & NA & D,UI\\ 
\{$y_{I_{11}}$, $y_{T_{11}}$, $y_{I_{BP1}}$, $y_{I_{BP2}}$\} & D,UI & D,UI & D,UI & D,UI & NA & D,UI & D,UI\\ 
\{$y_{V_{11}}$, $y_{T_{11}}$, $y_{I_{BP1}}$, $y_{I_{BP2}}$\}   & D,UI & D,UI & D,UI & NA & D,UI & D,UI & D,UI  \\ 
\midrule
\end{tabular}}
\end{table*}

Table \ref{t5} shows the fault detectability and isolability matrix for the single cell system with 4 sensors. With sensor set \{$y_{I_{11}},y_{V_{11}},y_{T_{11}},y_{I_{BP}}$\}, all faults can be detectable. The internal short circuit fault and all sensor faults can be uniquely isolable, while the  external short circuit fault and the load current sensor fault can be isolable from other faults but these two faults cannot be isolated from each other. From the table it can be seen that \{$y_{I_{11}},y_{V_{11}},y_{I_{BP1}},y_{I_{BP2}}$\}, \{$y_{I_{11}},y_{T_{11}},y_{I_{BP1}},y_{I_{BP2}}$\}, \{$y_{V_{11}},y_{T_{11}},y_{I_{BP1}},y_{I_{BP2}}$\} are minimal sensor sets achieving fault isolability. From the DM-decomposition result (choose sensor set \{$y_{I_{11}},y_{V_{11}},y_{I_{BP1}},y_{I_{BP2}}$\} as an example, shown in Fig. \ref{F12}(f)), it can be seen that all faults are in the over-determined part and each fault is in a unique equivalence class, which means every fault is uniquely isolable from other faults.

In this section, we have illustrated the fault detectability and isolability for a single cell system with all possible sensor combinations. In the following subsections, we focus on the minimal sensor set that can achieve complete fault isolability for a battery pack.

\subsection{Minimal sensor set for generalized mPnS topology}
\label{PS minimal}

\par
\begin{figure}[]
\centering
\includegraphics[width=9.5cm]{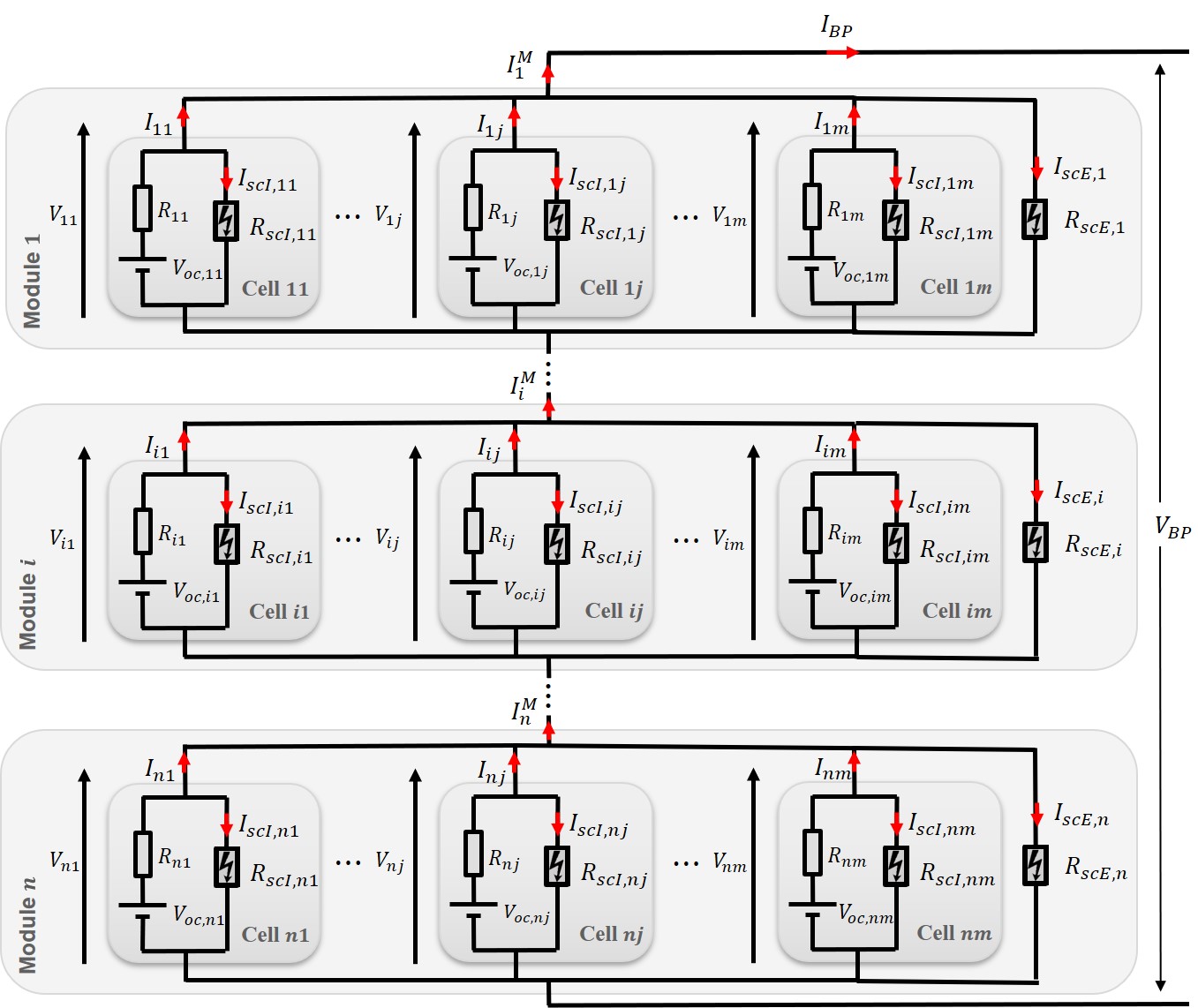}
\caption{Diagram of internal and external short circuit for $m$P$n$S battery pack topology}
\label{SP diagram with faults}

\centering
\includegraphics[width=9.5cm]{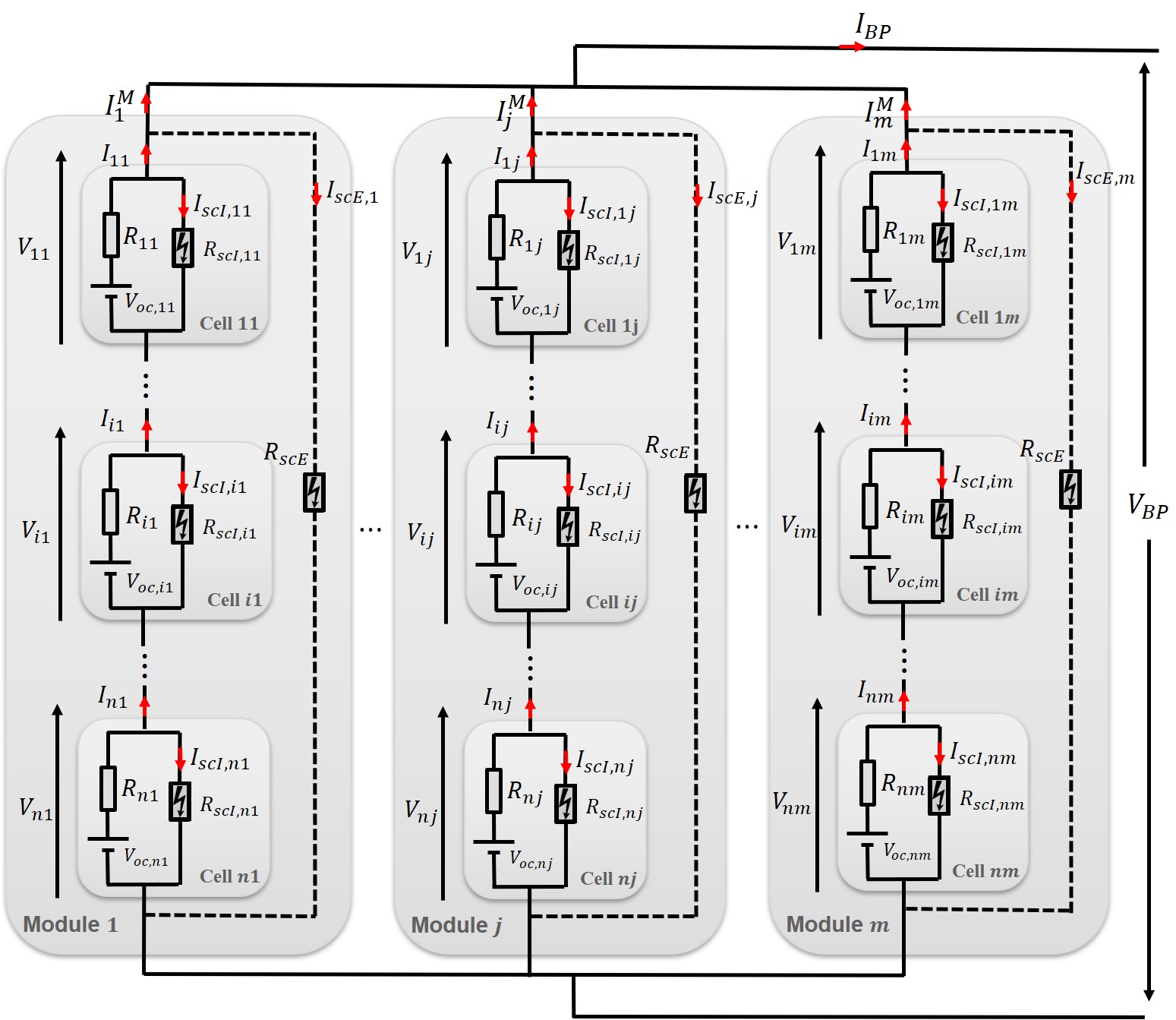}
\caption{Diagram of internal and external short circuit for $n$S$m$P battery pack topology}
\label{PS diagram with faults}
\end{figure}

 The generalized diagram including internal and external short circuit faults for the $m$P$n$S topology is shown in Figs. \ref{SP diagram with faults}. Note that, for the $m$P$n$S we only discuss cases when $m>1$, which means in each module there are at least two cells in parallel.  
In general, the set of sensor faults that needs to be diagnosed depends on the selected sensor set. Further, every single cell has the possibility of suffering from an internal short circuit. To represent internal short circuit faults at the cell level, every cell is modeled with an internal short circuit fault signal in it.  The set of internal short circuit faults we seek to diagnose is \{$f_{scI,11},\cdots,f_{scI,ij},\cdots,f_{scI,nm},$\}. 

\begin{table*}[t!]
	\caption{Summary of the minimal sensor set to achieve fault isolability for $m$P$n$S topology ($n>0,m>1$) ($Z$ represents current or temperature)}
 	\centering
 	\label{t6}
 	\scriptsize
 	\renewcommand{\arraystretch}{1.3}
\begin{tabular}{lllll}
\toprule
\multicolumn{1}{l}{}
              ${\rm{\#}}$ of module &Topology & Sensor set  & ${\rm{\#}}$ of sensors & ${\rm{\#}}$ of choices  \\ 
              \midrule
\multirow{4}{*}{$n$=1} &2P1S & \{$y_{I_{BP1}},y_{I_{BP2}},y_{Z_{11}},y_{Z_{12}}$\} & $2+1\times2$ & $2^2$  \\ 
&3P1S & \{$y_{I_{BP1}},y_{I_{BP2}},y_{Z_{11}},y_{Z_{12}},y_{Z_{13}}$\} &$2+1\times3$ &$2^3$  \\ 
             & \vdots & \vdots & \vdots &\vdots   \\ 
& $m$P1S & \{$y_{I_{BP1}},y_{I_{BP2}},y_{Z_{11}},\cdots,y_{Z_{1j}},\cdots,y_{Z_{1m}}$\} & $2+1\times m$& $2^m$  \\ \midrule
 \multirow{4}{*}{$n$=2} & 2P2S & \{$y_{I_{BP}},y_{Z_{11}},y_{Z_{12}},y_{Z_{21}},y_{Z_{22}}$\} & $1+2\times2$ &$2^{2m}$  \\ 
          & \vdots & \vdots & \vdots &\vdots  \\
& $m$P2S & \{$y_{I_{BP}},y_{Z_{11}},\cdots,y_{Z_{ij}},\cdots,y_{Z_{2m}}$\} & $1+2\times m$ &$2^{2m}$  \\ \midrule
\multirow{6}{*}{$n$\textgreater{}2} & 2P3S & \{$y_{Z_{11}},y_{Z_{12}},y_{Z_{21}},y_{Z_{22}},y_{Z_{31}}y_{Z_{32}}$\} & $3\times2$ &$2^{3m}$  \\ 
          & \vdots  &\vdots & \vdots &\vdots   \\ 
 & $m$P3S & \{$y_{Z_{11}},\cdots,y_{Z_{ij}},\cdots,y_{Z_{3m}}$\} & $3m$ & $2^{3m}$  \\
          & \vdots   & \vdots& \vdots &\vdots  \\ 
 & $m$P$n$S & \{$y_{Z_{11}},\cdots,y_{Z_{ij}},\cdots,y_{Z_{nm}}$\} & $nm$ & $2^{nm}$  \\ 
\bottomrule
\end{tabular}
\end{table*}

Every module has the possibility to suffer from external short circuit. The set of external short circuit faults that needs to be diagnosed is \{$f_{scE,1},f_{scE,2},\cdots,f_{scE,i}, \cdots,f_{scE,n},$\}. The possible sensor positions are \{$I_{BP},u_{11},\cdots,u_{ij},\cdots,u_{nm}$\}, where $u$ represents current, voltage or temperature of each cell. Table \ref{t6} lists the minimal sensor set to achieve fault isolability for $m$P$n$S topology obtained using the sensor placement toolbox. In order to uniquely isolate each fault, the sensor set installation needs to meet both of the following two requirements: 
\begin{enumerate}
    \item each cell should be equipped with a sensor $Z$ which can measure current or temperature;
    \item two sensors to measure the load current $I_{BP}$ when $n=1$; one sensor to measure the load current when $n=2$; no sensor is needed to measure the battery pack current when $n>2$.
\end{enumerate}
The number of load current sensor varies with $n$. This is true because as the battery pack scales up, the variable $I_{BP}$ will be contained in a greater number of equations (instances of KCL), see Appendix A.A. As a result, the redundancy of $I_{BP}$ increases automatically without the need of sensor. 
 

\subsection{Minimal sensor set for generalized nSmP topology}
\label{SP minimal}
The generalized diagram including internal and external short circuit faults for the $n$S$m$P topology is shown in Figs. \ref{PS diagram with faults}. Note that we only consider cases when $n>1$, which means in each module there are at least two cells in series. The set of sensor faults that needs to be diagnosed depends on the selected sensor set. Every single cell has the possibility of suffering from an internal short circuit. To represent internal short circuit faults at the cell level, every cell in this battery pack topology is modeled with an internal short circuit fault signal in it. The set of internal short circuit faults we seek to diagnose is \{$f_{scI,11},\cdots,f_{scI,ij},\cdots,f_{scI,nm},$\}. For $n$S$m$P topology, if more than one module suffers from an external short circuit, it is not possible to isolate these external short circuits from one other. So, when the sensor placement for fault isolability is performed, only one external short circuit is considered in the pack, and therefore only one external short circuit fault \{$f_{scE,j}$\} is to be diagnosed. If \textit{Module} $j$ is suffering from an external short circuit, all cells in this faulty module are shorted by this fault as shown in Fig. \ref{PS diagram with faults}. Since every module has the possibility of experiencing an external short circuit, we find the minimal sensor set that can achieve faults isolability regardless the location of external short circuit. For the $n$S$m$P topology, the possible sensor positions are \{$I_{BP},u_{11},\cdots,u_{ij},\cdots,u_{nm}$\}. $u$ represents current, voltage or temperature.
 
\begin{table*}[htb]
 	\caption{Summary of the minimal sensor set to achieve fault isolability for $n$S$m$P topology ($n>1,m>0$) ($Y$ represents voltage or temperature)}
 	\centering
 	\scriptsize
 	\label{t7}
 	\renewcommand{\arraystretch}{1.3}
\begin{tabular}{lllll}
\toprule 
\multicolumn{1}{l}{} ${\rm{\#}}$ of module& Topology & Sensor set  &${\rm{\#}}$ of sensors &${\rm{\#}}$ of choices   \\ 
              \midrule
\multirow{4}{*}{$m$=1} &  2S1P & \{$y_{I_{BP1}},y_{I_{BP2}},y_{Y_{11}},y_{Y_{21}}$\}  & $2 + 1 \times 2$ & $2^2$ \\
                &   3S1P & \{$y_{I_{BP1}},y_{I_{BP2}},y_{Y_{11}},y_{Y_{21}},y_{Y_{31}}$\}& $2 + 1 \times 3$ & $2^3$ \\
                  &  \vdots & \vdots & \vdots & \vdots \\
                  & $n$S1P & \{$y_{I_{BP1}},y_{I_{BP2}},y_{Y_{11}},\cdots,y_{Y_{i1}},\cdots,y_{Y_{n1}}$\} & $2 + 1 \times n$ & $2^n$ \\
                  \midrule
 \multicolumn{1}{l}{$m$\textgreater{}1} & $n$S$m$P & \begin{tabular}[c]{@{}l@{}}\{$y_{I_{BP1}},y_{I_{BP2}},\sum\limits_{j = 1}^m {{A_j}} $\}\\$B_{j}=\{Y_{ij}$\}, $i=1,\cdots,n$\\ $A_{j} \subseteq B_{j}$ and there are ($n-1$) elements in $A_{j}$ \end{tabular}& $2 + m \times(n-1)$ &  ${\left( {{2^{n - 1}} \times \left( {\begin{array}{*{20}{c}}
{n - 1}\\
n
\end{array}} \right)} \right)^m}$ \\

\bottomrule
\end{tabular}
\end{table*}

Table \ref{t7} lists the minimal sensor set to achieve fault isolability for $n$S$m$P topology regardless the location of external short circuit fault signal. To uniquely isolate each fault, the sensor set installation needs to meet the following two requirements at the same time:
\begin{enumerate}
    \item when $m=1$, each cell should be equipped with one sensor $Y$ which can measure voltage or temperature;  when $m>1$, in each module, $n-1$ cells should be equipped with a sensor $Y$ which can measure voltage or temperature. These $n-1$ cells can be chosen arbitrarily from the $n$ cells in each module;
    \item duplicate sensors to measure the load current $I_{BP}$. 
\end{enumerate}

\noindent As explained for the case of  $n$S$m$P topology, the need of multiple load current measurements depends on how many times the variable $I_{BP}$ appears in equations. Since in the equations of the $n$S$m$P topology, $I_{BP}$ appears only once (see Appendix A.B), two load current sensors are needed to achieve complete fault isolation. If two $n$S$m$P packs are connected in series, only one load current sensor is needed. If more than two $n$S$m$P packs are connected in series, a load current sensor is no longer necessary because of the redundancy of $I_{BP}$ intrinsically contained in the equations (instances of KCL).
\par In this section, the minimal sensor sets obtained using the sensor placement toolbox for two generalized battery pack topologies are illustrated. The minimal sensor set is the least among all sensor sets that can provide the system with a sufficient degree of AR to isolate every fault. A sufficient degree of AR permits finding different MSO subsystems, and each MSO subsystem can then generate a residual after choosing a proper matching, as illustrated in Section \ref{sec structual analysis}. Each residual is sensitive to a certain set of faults. With the fault signature matrix, solutions can be found to make every fault uniquely isolable from other faults.
\par Note that the process of finding the minimal sensor set using sensor placement toolbox doesn't consider whether there is a loop or not that needs to be solved to compute the unknown variables and generate the residuals in a matching (if no loop is present, then the solution can be derived from a  alternated chain). The presence of loops will increase the computational effort and may limit the implementation of FDI scheme in practical application. Thus, we define an optimal sensor set to be one that satisfies the isolability requirement and that has minimal computational effort.
If a system is equipped with the minimal sensor set and there is not any computational loops in the residual generation, then, this minimal sensor set is also the optimal one. Otherwise, additional sensors are needed to form the optimal sensor set that results in no computational loops. The optimal sensor sets for two generalized battery pack topologies are discussed in Section \ref{optimal sensor set}

\section{Final Comments and Remarks}
\label{sec comments and remarks}

The methodology for battery pack fault diagnosis illustrated in this paper is based on understanding and exploiting the analytical redundancy in the system. The analytical redundancy required for fault diagnosis is in part inherently present in the analytical equations of the system, and in part added by installing sensors, which, in the context of structural analysis, convert unknown variables into known variables and help us determine which variables play a key role in diagnosing the faults. 
The minimal sensor set to achieve internal short circuit fault isolation is an intrinsic characteristics of  each topology. For the  $m$P$n$S topology, the cells in each module are in parallel so they share the same voltage. Thus only by adding current or temperature sensor can the fault isolation at the cell level be achieved. For $n$S$m$P topology, the cells in each module are in series so there is only one current. Thus, only voltage or temperature sensors can add redundancy to permit fault isolation at the cell level. This duality is a natural consequence of series vs. parallel circuits. On the other hand, temperature sensors can be effective in both topologies, as they are in principle sensitive to the heat generation caused by an internal short circuit.  On the contrary, in  our models, the external short circuit does not play a role in the heat balance equation. While temperature sensors could in principle be very useful, it is not practical to install temperature sensors in each cell due to: i) their slow dynamic response; ii) their cost; and iii) the difficulty in physically mounting the sensors at the manufacturing stage.
From the perspective of practical application, when adding a current or voltage sensor can provide the same FDI capability to a battery as adding a temperature sensor, current or voltage sensor is a preferred choice.

\subsection{Optimal sensor set for FDI}
\label{optimal sensor set}

\par According to the structural analysis for a single cell with different measurements illustrated in Section \ref{subsec singlecell diffsen}, see Fig. \ref{different sensing}, the addition of a voltage or a temperature sensor at the cell level will create computational loops. Only by adding a current sensor is it possible to express the solution in the form of alternated chains that allow solving for the unknown variables without the need to solve simultaneous equations (loops), and therefore resulting in lower computational effort.  Thus, the minimal sensor set for each battery pack topology for FDI obtained in \ref{PS minimal} and \ref{SP minimal} is further optimized by considering both FDI capability and computational effort.
\par \noindent The optimal sensor set for $m$P$n$S battery pack topology is listed below: 
\begin{enumerate}
    \item each cell should be equipped with a current sensor, preferred to  a temperature sensor which would result in computational loops in residual generation and which also has some practical limitations as explained earlier;
    \item two sensors to measure the load current $I_{BP}$ when $n=1$; one sensor to measure the load current when $n=2$; no sensor is needed to measure the battery pack current when $n>2$.
\end{enumerate}
The optimal sensor set for $n$S$m$P battery pack topology is:
\begin{enumerate}
    \item when $m=1$, each cell should be equipped with a voltage sensor;  when $m>1$, in each module, $n-1$ cells should be equipped with a voltage sensor. These $n-1$ cells can be chosen arbitrarily from the $n$ cells in each module; While in theory a temperature sensor could be used in place of a voltage sensor, in the optimal set we remove this option for the practical reasons stated earlier. 
    \item each module should be equipped with a current sensor since when each cell's current is known, computational loops are avoided in  residual generation. \item duplicate sensors to measure the load current $I_{BP}$.
\end{enumerate}
\subsection{Traditional sensor set evaluation}
Today, the most common sensor set for battery packs used in automotive applications includes \cite{rahimi2013battery}: 
\begin{enumerate} \item a load current sensor to measure $I_{BP}$; \item voltage sensors for each cell to permit voltage balancing and over-charge/over-discharge protection functions in the BMS (in the $m$P$n$S topology cells that are in parallel share a single voltage sensor, in the $n$S$m$P topology each cell has its own voltage sensor); \item a temperature sensor per module.\end{enumerate} In this paper, we refer to this sensor set as the traditional one. To evaluate the diagnosability for traditional sensor set, thermal model at module level is needed to provide a module temperature variable ($T^{M}$) to permit place a temperature sensor per module for monitoring any over-temperature conditions and thus correctly managing the cooling system.. 

\noindent Thermal model for \textit{Module} $i$ in $m$P$n$S topology:
\begin{eqnarray}
\label{module_thermal_SP}
{T_{i}^{M}}=\frac{1}{m}\sum\limits_{j=1}^m {{k_{ij}}{T_{ij}}}
\end{eqnarray}
Thermal model for \textit{Module} $j$ in $n$S$m$P topology:
\begin{eqnarray}
\label{module_thermal_PS}
{T_{j}^{M}}=\frac{1}{n}\sum\limits_{i=1}^n {{k_{ij}}{T_{ij}}}
\end{eqnarray}
Where $k_{ij}$ represents the weighted average temperature of the module and it depends on the distance between the $ij$th cell and the temperature sensor. This model assumes that thermal connections among cells follow the same architecture of the electrical connections within the module, and that there is no thermal interaction between modules. 

\begin{figure}[]
\centering
\includegraphics[width=8.5 cm]{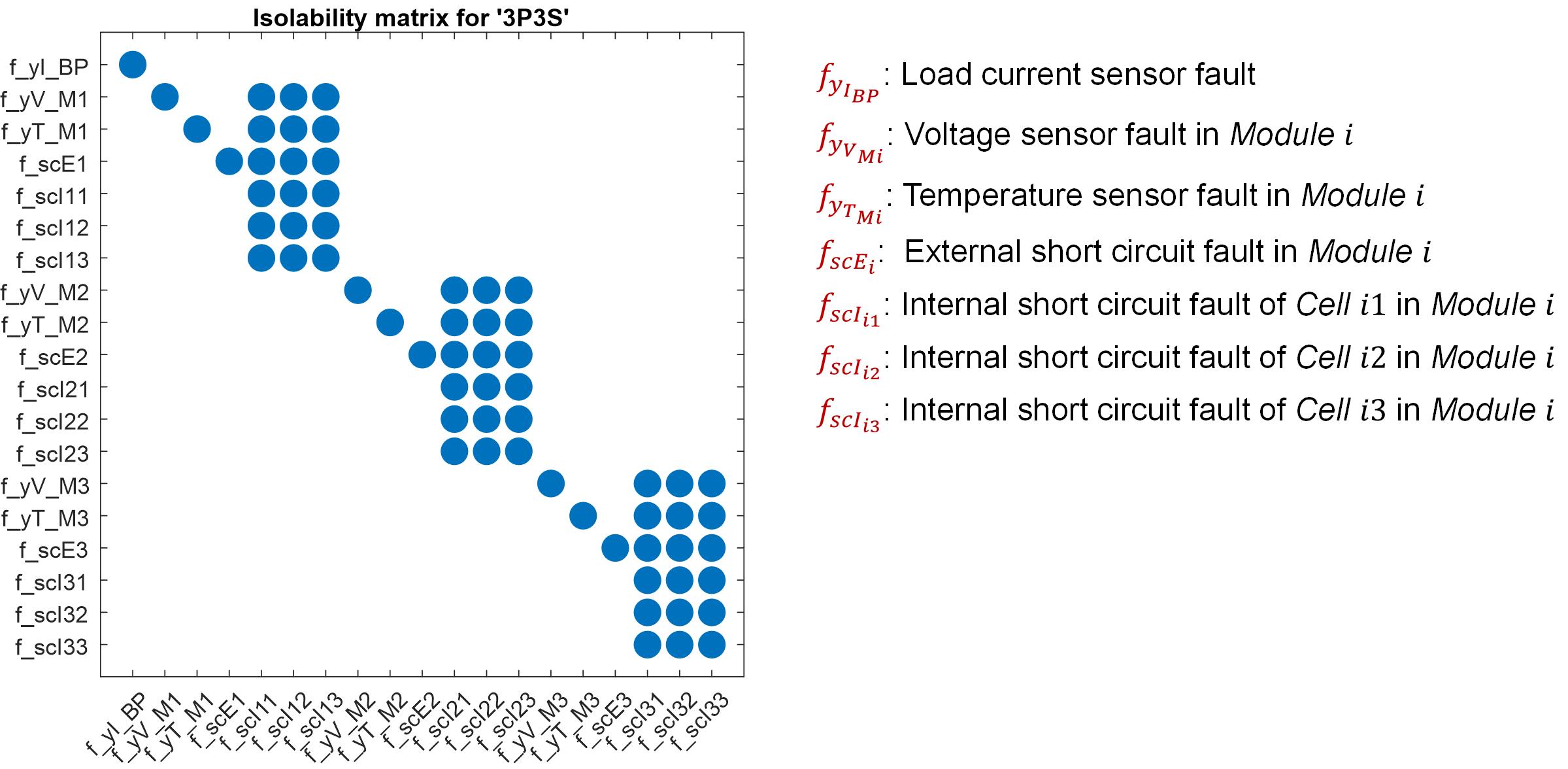}
\caption{Fault isolability matrix of 3P3S topology battery pack with traditional sensor set \{$y_{I_{BP}},y_{V_{1}^M},y_{V_{2}^M},y_{V_{3}^M},y_{T_{1}^M},y_{T_{2}^M},y_{T_{3}^M}$\}}
\label{3P3S_traditional}
\end{figure}

An $m$P$n$S battery pack with a traditional sensor set has the ability to detect all faults. As for isolability, a traditional sensor set can isolate faults in a module from faults in another module while it fails to isolate every fault within the module (at the cell level). For example, Fig. \ref{3P3S_traditional} shows the fault isolability matrix of a 3P3S topology battery pack with a traditional sensor set. It can be seen that the battery pack current sensor fault can be uniquely isolated from the other faults, while faults in \textit{Modules} 1, 2 and 3 are isolable from each other but faults within the module cannot be uniquely isolated. In each module, module voltage sensor fault, module temperature sensor fault and external short circuit fault are isolated from each other, while they are all not isolated from the three internal short circuit. The three internal short circuit faults are not isolable from each other. If the 3P3S topology battery pack were equipped with the optimal sensor set \{$y_{I_{11}},y_{I_{12}},y_{I_{13}},y_{I_{21}},y_{I_{22}},y_{I_{23}},y_{I_{31}},y_{I_{32}},y_{I_{33}}$\} resulting from the analysis done in this paper, all faults could be uniquely isolated from each other. However, the optimal sensor set does not include voltage sensors, which suggests that in a $m$P$n$S battery pack equipped with only the optimal sensor set for FDI, the BMS would not be able to perform voltage balancing. Likewise, a $m$P$n$S battery pack equipped with only a voltage sensor in each module for balancing would not achieve FDI.


\begin{figure}[htb]
\centering
\includegraphics[width=8.5 cm]{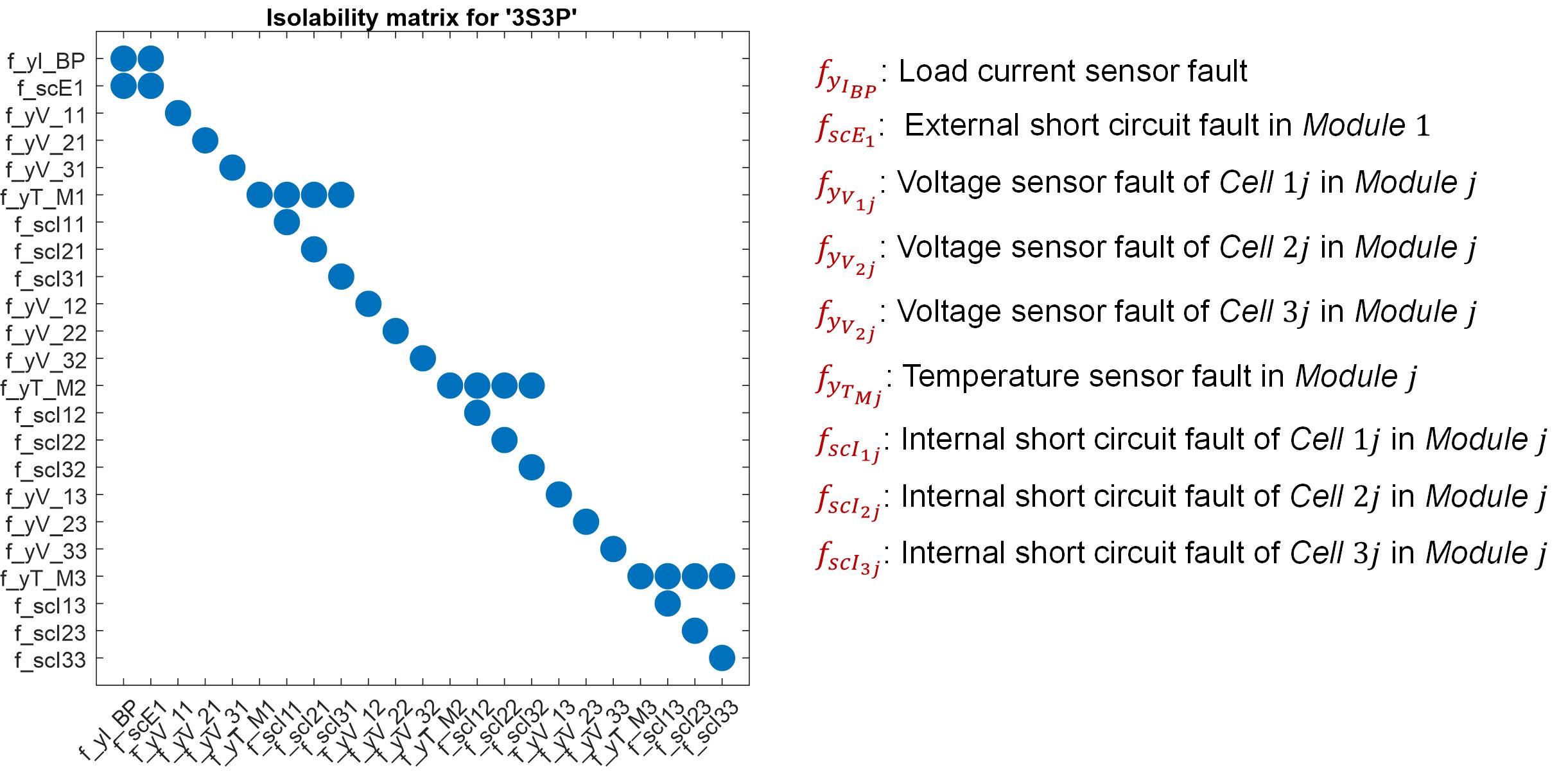}
\caption{Fault isolability matrix of 3S3P topology battery pack with traditional sensor set \{$y_{I_{BP}},y_{T_{1}^M},y_{T_{2}^M},y_{T_{3}^M},y_{V_{11}},y_{V_{21}},y_{V_{31}},y_{V_{12}},y_{V_{22}},y_{V_{32}},\\y_{V_{13}},y_{V_{23}},y_{V_{33}}$\}}
\label{3S3P_traditional}
\end{figure}

For a $n$S$m$P topology battery pack with the traditional sensor set, the faults in each cell can be uniquely isolated. The battery pack current sensor fault and the external short circuit fault cannot be isolated from each other, but can be isolated from the other faults. In each module, the temperature sensor fault is not isolable from the internal short circuit faults. Similarly, 3S3P is chosen as an example and assume that the external short circuit is in \textit{Module} 1. Fig. \ref{3S3P_traditional} shows the fault isolability matrix of 3S3P topology with the traditional sensor set. It can be seen that the load current sensor fault and external short circuit fault fall in the same equivalent class and cannot be isolated from each other. Internal short circuit fault in each cell are isolated from each other while in each module, the module temperature sensor fault is not isolable from the three internal short circuit faults. If the 3S3P topology battery pack were equipped with the optimal sensor set \{$y_{I_{BP1}},y_{I_{BP2}},y_{I_{1}^M},y_{I_{2}^M},y_{I_{3}^M},y_{V_{11}},y_{V_{12}}, y_{V_{21}},y_{V_{22}}, y_{V_{31}},y_{I_{32}}$\} derived in this paper, all faults could be uniquely isolated from each other.  Note that in the optimal sensor set for $n$S$m$P topology, although a current sensor needs to be added in each module, two instead of three voltage sensors are sufficient for each module (series string) to achieve fault isolation, no matter in which module the external short circuit fault occurs. This indicates that for $n$S$m$P battery pack topology, the voltage sensor that has been already installed in each cell for balancing can be also used as a part of the optimal sensor set for FDI.

Finally, it should be pointed out that as long as the sensor set selected in a pack design includes as a subset the optimal diagnostic sensor set for isolability derived in this paper, multiple design objectives can be met.
\section{Conclusion}
\label{sec concl}
The work presented in this paper uses the tools of structural analysis for diagnosis to derive some fundamental characteristics of two principal battery pack topologies from a diagnostic perspective. The equivalent circuit models and lumped-parameter thermal models used to represent each cell permit the determination of the analytical redundancy that is intrinsic in the battery system (always -1 regardless of pack topology and number of cells). The methods developed in this work are first applied to the simplest representation (a single cell) to illustrate how one can select a minimal/optimal sensor set to achieve detectability and isolability of faults, and are then generalized to the $m$P$n$S and $n$S$m$P topologies to yield results that are generally applicable to either topology regardless of cell number. Further, the model and methods applied to a single cell can be applied in exactly the same way at the module level, regardless of the module internal configuration, thus making this approach completely scalable - a property that is very important when one considers applications with hundreds or thousands of individual cells, such as in automotive, aerospace and grid support applications. For future work, we are interested in exploring a distributed FDI scheme for a battery pack installed with an optimal sensor set presented in this work.


%

\appendices
\section{}

\subsection{Model for mPnS topology battery pack with faults:}

$n>0; m>1; i=1,\cdots,n; j=2,\cdots,m$

${e_1}:{V_{ij}} = {V_{oc,ij}} - {R_{ij}}(I_{ij}+I_{scI,ij})$ 

$e_{2}: \frac{dSoC_{ij}}{dt}=-\frac{(I_{ij}+I_{scI,ij})}{Q_{ij}}$

${e_{3}}:{V_{oc,ij}} = f(SoC_{ij})$

${e_{4}}:m{c_p}\frac{{d{T_{ij}}}}{{dt}} = {R_{ij}}{\left( {(I_{ij}+I_{scI,ij})} \right)^2} - {Q_{TMS_{ij}}}$

${e_{5}}:{I_{scI,ij}} = \left( {\frac{V_{ij}}{{{R_{scI}}}}} \right){f_{scI,ij}}$

${e_{6}}:{I_{scE,i}} = \left( {\frac{V_{i1}}{{{R_{scE}}}}} \right){f_{scE,i}}$

${e_{7}}:{I_{i1}}+{I_{i2}}+\cdots+{I_{ij}} = {I_{BP}} + {I_{scE,i}}$

${e_{8}}:{V_{i1}} = {V_{i2}}=\cdots= {V_{ij}}$

\subsection{model for nSmP topology battery pack with faults:}

$m>0; n>1; i=2,\cdots,n; j=1,\cdots,m$
\par

${e_1}:{V_{ij}} = {V_{oc,ij}} - {R_{ij}}(I_{ij}+I_{scI,ij})$ 

$e_{2}: \frac{dSoC_{ij}}{dt}=-\frac{(I_{ij}+I_{scI,ij})}{Q_{ij}}$

${e_{3}}:{V_{oc,ij}} = f(SoC_{ij})$

${e_{4}}:m{c_p}\frac{{d{T_{ij}}}}{{dt}} = {R_{ij}}{\left( {(I_{ij}+I_{scI,ij})} \right)^2} - {Q_{TMS_{ij}}}$

${e_{5}}:{I_{scI,ij}} = \left( {\frac{V_{ij}}{{{R_{scI}}}}} \right){f_{scI,ij}}$

${e_{6}}:{V_{1j}}+\cdots+{V_{ij}}+\cdots+{V_{nj}} = {V_{j}^{M}}$

${e_{7}}:{I_{1j}} =\cdots= {I_{ij}}=\cdots={I_{nj}}= {I_{j}^{M}}$
\par
${e_{8}}:{I_{1}^{M}}+\cdots+{I_{j}^{M}}+\cdots+{I_{m}^{M}} = {I_{BP}}$

${e_{9}}:{V_{1}^{M}} =\cdots= {V_{j}^{M}}=\cdots= {V_{m}^{M}}={V_{BP}}$

If \textit{Module} $j$ suffers from the external short circuit as shown in Fig.\ref{PS diagram with faults}, $e_{7}$ should be substituted by $e_{10}$ and $e_{11}$.

${e_{10}}:{I_{scE,j}} = \left( {\frac{V_{1j}+\cdots+V_{ij}+\cdots+V_{nj}}{{{R_{scE}}}}} \right){f_{scE,j}}$

${e_{11}}:{I_{1j}} =\cdots= {I_{ij}}=\cdots= {I_{nj}}={I_{j}^{M}} + {I_{scE,j}}$

\ifCLASSOPTIONcaptionsoff
  \newpage
\fi



%
\bibliographystyle{IEEEtran}
\bibliography{IEEEabrv,aaa}

%

\end{document}